\documentclass[final,5p,times,twocolumn]{elsarticle}

\usepackage{graphics}
\usepackage{multirow}
\usepackage[usenames]{color}

\usepackage[dvipsnames]{xcolor}

\usepackage{amssymb}
\usepackage{amsmath}
\usepackage{tabularx}

\journal{Journal of Petroleum Science and Engineering}

\begin{document}
\newcolumntype{L}[1]{>{\raggedright\arraybackslash}p{#1}}
\newcolumntype{C}[1]{>{\centering\arraybackslash}p{#1}}
\newcolumntype{R}[1]{>{\raggedleft\arraybackslash}p{#1}}
\begin{frontmatter}

\title{A Predictive Model for Steady-State Multiphase Pipe Flow: Machine Learning on Lab Data}


\author[SKT]{E.A.~Kanin}
\author[SKT]{A.A.~Osiptsov}
\ead{a.osiptsov@skoltech.ru}
\author[SKT]{A.L.~Vainshtein}
\author[SKT]{E.V.~Burnaev}

\address[SKT]{Skolkovo Institute of Science and Technology (Skoltech), 3 Nobel Street, 143026, Moscow, Russian Federation}

\begin{abstract}
\hspace{0.25cm}

Engineering simulators used for steady-state multiphase flows in oil and gas wells and pipelines are commonly utilized to predict pressure drop and phase velocities. Such simulators are typically based on either empirical correlations (e.g., Beggs and Brill, Mukherjee and Brill, Duns and Ros) or first-principles mechanistic models (e.g., Ansari, Xiao, TUFFP Unified, Leda Flow Point model, OLGAS). The simulators allow one to evaluate the pressure drop in a multiphase pipe flow with acceptable accuracy. However, the only shortcoming of these correlations and mechanistic models is their applicability (besides steady-state versions of transient simulators such as Leda Flow and OLGA). Empirical correlations are commonly applicable in their respective ranges of data fitting; and mechanistic models are limited by the applicability of the empirically based closure relations that are a part of such models. In order to extend the applicability and the accuracy of the existing accessible methods, a method of pressure drop calculation in the pipeline is proposed. The method is based on well segmentation and calculation of the pressure gradient in each segment using three surrogate models based on Machine Learning (ML) algorithms trained on a representative lab data set from the open literature. The first model predicts the value of a liquid holdup in the segment, the second one determines the flow pattern, and the third one is used to estimate the pressure gradient. To build these models, several ML algorithms are trained such as Random Forest, Gradient Boosting Decision Trees, Support Vector Machine, and Artificial Neural Network, and their predictive abilities are cross-compared. The proposed method for pressure gradient calculation yields $R^2 = 0.95$ by using the Gradient Boosting algorithm as compared with $R^2 = 0.92$ in case of Mukherjee and Brill correlation and $R^2 = 0.91$ when a combination of Ansari and Xiao mechanistic models is utilized. The application of the above-mentioned ML algorithms and the larger database used for their training will allow extending the proposed methodology to a wider applicability range of input parameters as compared to standard accessible techniques. The method for pressure drop prediction based on ML algorithms trained on lab data is also validated on three real field cases. Validation indicates that the proposed model yields the following coefficients of determination: $R^2 = 0.806, 0.815$ and 0.99 as compared with the highest values obtained by commonly used techniques: $R^2 = 0.82$ (Beggs and Brill correlation), $R^2 = 0.823$ (Mukherjee and Brill correlation) and $R^2 = 0.98$ (Beggs and Brill correlation). Hence, the method for calculating the pressure distribution could give comparable or even higher scores on field data by contrast to correlations and mechanistic models. This fact is an indicator that the model can be scalable from the lab to the field conditions without any additional retraining of ML algorithms.
\end{abstract}

\end{frontmatter}

\section{Introduction}
\label{Intro}
\subsection{Brief introduction and literature overview}
\label{Introduction}
Multiphase flows in pipes of a circular cross-section are encountered in many industrial applications, such as drilling of oil wells~\cite{ladva2000multiphase}, \cite{sun2018multiphase}, multistage fracturing and cleanup of oil or gas wells~\cite{osiptsov2017review}, transport of hydrocarbons over surface~\cite{bratland2010pipe}. Multiphase flow is understood as a simultaneous flow of a mixture of two or more phases (several phases such as liquid, gas or solid). The flow at each pipe cross-section is characterized by the volume fractions of the phases, pressure (which is typically assumed the same in all phases in a given cross-section), and the flow pattern according to the physical distribution of phases. The volume fraction of the liquid is widely attributed to as the liquid holdup. During multiphase flow, the flow regime depends on the magnitudes of forces that act on the fluid from other fluids or the pipe wall. In turn, the local pressure gradient significantly depends on the flow pattern. Hence, in order to determine the pressure gradient (and then the global pressure drop), one needs to determine the volume fractions and the flow regime along the pipeline. There is numerous literature on multiphase flow modelling in pipelines (see the review in~\cite{bratland2010pipe}), while less effort is devoted to the application of Machine Learning (for brevity, we will use the acronym ML in what follows) algorithms in the identification of flow patterns. In these papers, authors plotted experimental data points in suitable coordinates and tried to construct models in order to match these points. In paper \cite{osman2004artificial} author created Artificial Neural Network (ANN), in \cite{li2014multiphase} the authors applied Support Vector Machine algorithm (SVM), and in \cite{popa2015fuzzynistic} fuzzy inference system was used. 

Accurately determining the liquid holdup is also important in planning the design of separation equipment. For example, the slug flow regime can be dangerous for the separator, when a significant liquid mass comes unexpectedly to surface from a long near-horizontal wellbore. Several papers were devoted to determining this parameter by machine learning tools. For example, in \cite{osman2004artificial} and \cite{shippen2002neural} authors have applied ANN. 

In order to calculate pressure distribution in a pipe, segmentation and numerical algorithm are used. During multiphase flow in tube flow pattern and pressure gradient change along the pipe. Therefore, to solve this problem, it is necessary to break down the whole pipe into segments. Within each segment, the flow regime can be considered homogeneous, and the pressure gradient is approximately constant. Each step of the numerical algorithm calculates the pressure gradient within the segment. Multiphase flow correlations and mechanistic models are commonly used for this purpose. 

Correlations were developed upon laboratory experiments. Among the most widely used are Aziz and Govier \cite{aziz1972pressure}, Beggs \& Brill \cite{beggs1973study}, Mukherjee \& Brill \cite{mukherjee1983liquid}, and others. Many articles are concerned with issues of limits to the applicability of multiphase flow correlations. Authors of these papers conclude that each correlation could be applied only in its range of input parameters for obtaining results with acceptable accuracy.
There are also several mechanistic models with semi-empirical closure relations, which are used to predict different multiphase flow characteristics. The most popular ones are Hasan and Kabir \cite{hasan1986study}, Ansari \cite{ansari1990comprehensive}, TUFFP Unified \cite{zhang2003unified_1}, \cite{zhang2003unified_2} and others. 

Furthermore, there are also two steady-state multiphase flow models that have a relatively high accuracy of pressure drop calculations, namely, Leda Flow Point model \cite{danielson2005leda} and OLGAS \cite{bendiksen1991dynamic}. These mechanistic models are steady-state versions of corresponding transient simulators.

The literature review shows that researchers applied ML algorithms for pressure gradient prediction and direct output pressure estimation. In the article \cite{osman2005artificial} authors predicted the bottomhole pressure using ANN. In contrast, in the paper \cite{li2014combined} authors forecasted value of bottomhole pressure via ANN, but they suggested to use segmentation of the well and identify flow regime in each segment. 

The present work is a continuation and extension of~\cite{kanin2018method}. Compared to the earlier work, all the results are new. We expanded the database by more than 20\%, changed the set of input parameters used for constructing the surrogate models, and modified the resulting sub-model for pressure gradient calculation. Now we use the momentum conservation equation as a framework and have a separate sub-model for the friction factor coefficient. The workflow is aimed at the prediction of the pressure distribution. A sensitivity study is carried out. Finally, the model trained on lab data is applied to the field cases without any retraining.


\subsection{Precise problem formulation}

We consider the class of steady-state multiphase gas-liquid flows in wells and pipelines of circular cross-section and an arbitrary inclination angle to the horizontal, in the gravity field. In oil production, this class of flows can be encountered at various stages of the well life, from cleanup after drilling, through cleanup after fracturing to the production stage. The key to control oilfield services operations is the ability to accurately predict the pressure drop along the well, while typical measurements of pressure are taken on surface (with the memory pressure gauges installed downhole, but the readings are rarely available in real-time, so the tool to calculate the pressure drop from surface to the bottom hole is required). Another issue is the pressure drop along the horizontal section of a (generally, multistage fractured) well. During flowback and production, the well is controlled by the choke on the surface and by the ESP (if the one is installed), but no measurements are taken in the horizontal section. Excessive drawdown (flowing the well at excessively high flow rates) may be dangerous to the well-fracture system, resulting in undesired geomechanics events, such as proppant flowback, tensile rock failure and fracture pinch-out in the near wellbore~\cite{osiptsov2017review}.

The new approach for pressure gradient calculation on the arbitrarily oriented pipe segment is proposed. The method is based on machine learning algorithms. The considered algorithms are trained on lab data set that is collected from open source: articles and dissertations. The new method consists of three surrogate models nested within each other (see a general discussion of surrogate modeling in \cite{GTApprox2016}). The first model predicts liquid holdup parameter. The second one identifies the flow regime in the segment. The value of liquid holdup is included in the input parameters of this model. The resultant model estimates the pressure gradient. Among input parameters of this model, flow pattern and value of liquid holdup are presented. 

This comprehensive approach of pressure gradient calculation, which includes information about flow pattern and liquid holdup, and surrogate models, trained on lab data from different sources, yields output value with high accuracy and allow obtaining the model with wider scope of applicability in comparison with standard techniques.


\section{Modeling approach}
\subsection{Details of modeling approach}
\begin{figure}[!htb]
	\centering
	\includegraphics[width = 0.5 \textwidth]{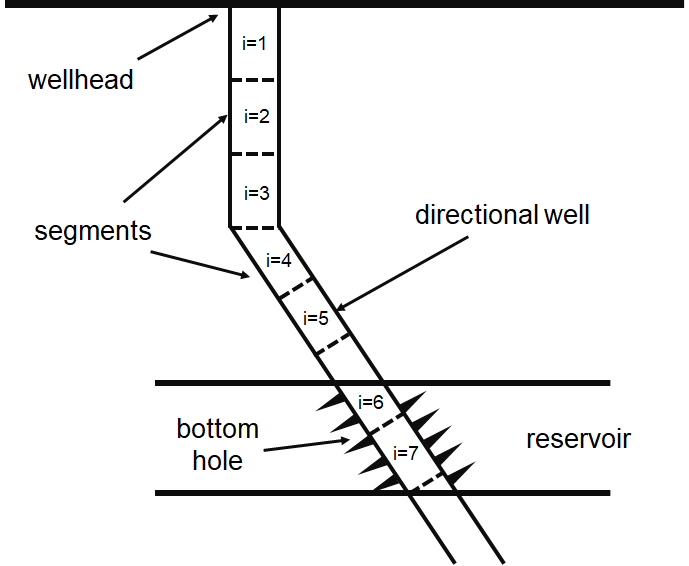}
	\caption{Schematic picture of the segmented well}\label{fig16}
\end{figure}

This part of the article is devoted to the description of the method of calculating the pressure distribution along a pipe. In open literature (e.g., \cite{brill1999multiphase}) this method is called the marching algorithm. It is the numerical algorithm for computation the integral of pressure gradient function along the pipe path: 
\begin{equation}
\Delta p = \int_{0}^{L}\frac{dp}{dL}(l')dl'.
\end{equation}
To perform the integration, the pipe should be divided into \textit{n} segments of length $\Delta \textit{L}_i$. Thus inside each segment, the flow type is approximately homogeneous, and the pressure gradient along the segment is approximately constant.
In Fig.~\ref{fig16}, an example of the segmented inclined well is represented.

In other words, the integrand function $\left(\frac{dp}{dL}\right) (l')$ is represented by a piece-wise constant approximation and the integral is transformed into summation:
\begin{equation}
\Delta p = \sum_{i = 1}^{n}\left(\frac{dp}{dL}\right)_i\Delta L_i,
\end{equation}
where $\Delta L_i$ is the length of $i$-th segment and $\left(\frac{dp}{dL}\right)_i$ is the pressure gradient along $i$-th segment of the pipe.

\begin{figure}[!htb]
	\centering
	\includegraphics[width = 0.45\textwidth]{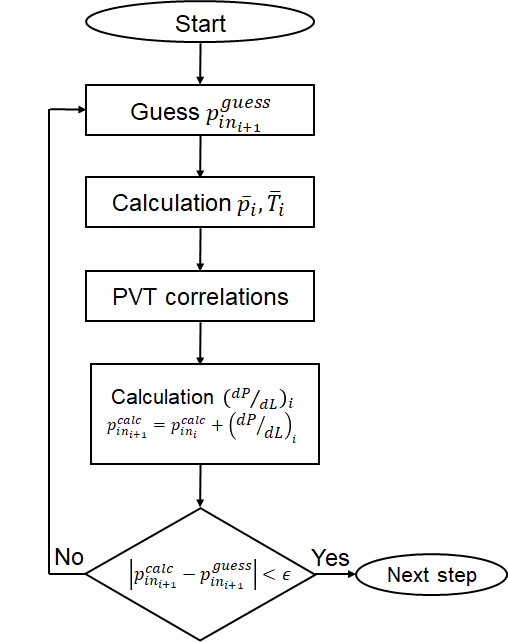}
	\caption{The flowchart of the marching algorithm.}\label{fig26}
\end{figure}

We show the flowchart of the marching algorithm in Fig.~\ref{fig26}.

Before calculation process the following parameters are known: pressure and temperature at the inlet $p_{in} = p_{in_1}$ and $T_{in}$, temperature at outlet $T_{out}$, liquid flow rate $Q_{l_{SC}}$, densities of oil, gas and water at standard conditions, gas-oil ratio (GOR), water cut (WC) and other parameters. In the numerical algorithm heat processes are not accurately considered that is why linear interpolation for temperature between inlet and outlet is used: $$T_i = T_{in} + \frac{T_{out} - T_{in}}{n} \cdot (i-1).$$
To calculate the flow parameters in each segment, the set of PVT correlations is utilized in which pressure and temperature are taken in the middle of the segment. 

\begin{figure}[!htb]
	\centering
	\includegraphics[width = 0.5 \textwidth]{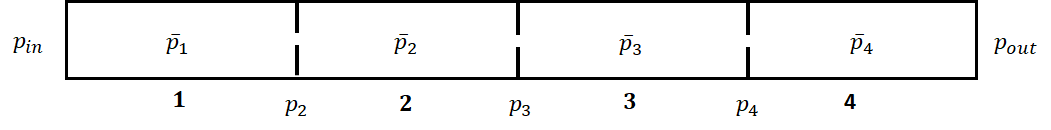}
	\caption{Schematic description of the iterative algorithm for pressure drop calculations}\label{fig1}
\end{figure}

To begin with, the first segment is considered. In order to calculate pressure at the inlet of the second segment $p_{in_2}$ (it is equal to the pressure at the outlet of the first segment), the iterative algorithm is applied. In each step of this algorithm, the value of outlet pressure of the first segment $p^{guess}_{in_2}$ is guessed and the pressure at the center of the segment $$\bar{p}_1 = \frac{p^{guess}_{in_2} - p_{in_1}}{2}$$ is calculated. Using $\bar{p}_1$, temperature at the center of the segment from linear interpolation and PVT correlation, pressure gradient $\left(\frac{dp}{dL}\right)_1$ are estimated and, consequently, pressure $$p^{calc}_{in_2} = p_{in_1} + \left(\frac{dp}{dL}\right)_1 \Delta L_1$$ is obtained. These calculations are continued until the execution of the following condition:
$$\left|p^{calc}_{in_2} - p^{guess}_{in_2}\right| < \epsilon,$$ 
where $\epsilon$ is desirable accuracy.
As a result, the input pressure of the second segment $$p_{in_{2}} = p_{in_1} + \left(\frac{dp}{dL}\right)_1 \Delta L_1$$ is derived. Continuing this iterative process, one could compute the output pressure: 
\begin{equation}
p_{out} = p_{in} + \sum_{i=1}^{n}\left(\frac{dp}{dL}\right)_i \Delta L_i.
\label{eq1}
\end{equation}
The schematic description of these calculations is shown in Fig.~\ref{fig1}.

At each step of the marching algorithm, it is necessary to compute the value of pressure gradient in the considered segment. For this purpose, the new approach is proposed. It is based on machine learning algorithms and consists of three surrogate models. The resultant model estimates value of the pressure gradient in the segment. Among input parameters of this model we also include the value of liquid holdup and flow pattern (besides velocities and PVT properties of liquid and gas at flowing conditions). These parameters are determined by the other two surrogate models. The next subsection of the article will be devoted to the detailed description of the proposed method of pressure gradient calculation. 
\subsection{Method of calculating the pressure gradient in the pipe segment}
\label{input parameters}
This subsection is devoted to the description of the proposed method for calculating the pressure gradient in the segment which could be oriented at an arbitrary angle to the horizontal direction (from $-90^{\circ}$ to $90^{\circ}$). Inclination angle equal to zero corresponds to a horizontal flow. Positive and negative angles correspond to uphill and downhill flows, respectively. 

The method of calculating the pressure gradient in the segment consists of three surrogate models nested within each other that are based on machine learning algorithms. The first model predicts the value of liquid holdup parameter in the pipe segment. The second model identifies the flow regime. It is necessary to highlight that among input parameters of this model the output result of the first model is used. The final model is targeted to the estimation of the pressure gradient. In this model, both output results of the first model and the second models are utilized.

For creating the surrogate models applicable for different types of liquids and gases and for reducing the number of input features when training ML algorithms, the following set of dimensionless variables was chosen according to the article~\cite{duns1963vertical}. These parameters are called velocity number of gas, velocity number of liquid, viscosity number and defined by the following equations:
\begin{equation}
N_{vg}=v_{sg}\sqrt[4]{\frac{\rho_l}{g \sigma_{lg}}},N_{vl}=v_{sl}\sqrt[4]{\frac{\rho_l}{g \sigma_{lg}}}, N_{\mu} = \mu_l \sqrt[4]{\frac{g}{\rho_l \sigma_{lg}^3}},
\label{eq2}
\end{equation}
where $v_{sg}$ and $v_{sl}$ are gas and liquid superficial velocities respectively, $\rho_l$ is a liquid density and $\sigma_{lg}$ is a surface tension between liquid and gas phases, $\mu_l$ is a liquid viscosity and $g$ is the gravitational acceleration.

In this set of dimensionless parameters, the diameter number $N_{d} = d \sqrt{\frac{\rho_l g}{\sigma_{lg}}}$ ($d$ is the tube diameter) is excluded because of its relatively narrow range of values in the case of lab data. This range almost does not overlap with the appropriate one in the case of field data (the comparison will be represented in the Section \ref{Case_study_section}). As a result, the surrogate model, trained on lab data, with the diameter number among input features could not be potentially applied for calculation connected with field data.

The next input parameter in the surrogate models is the no-slip liquid holdup defined by the following equation:
$$\lambda_l = \frac{v_{sl}}{v_m}$$
where $v_m = v_{sg} + v_{sl}$ is the mixture velocity.

Finally, we also include Reynolds ($\textrm{Re}$) and Froude ($\textrm{Fr}$) numbers
into input features. They are defined according to the equations:
$$\mathrm{Re} = \frac{\rho_{ns} v_m d}{\mu_{ns}}, \mathrm{Fr} = \frac{v_m^2}{gd}$$
where $\rho_{ns} = \rho_l \lambda_l + \rho_g (1 - \lambda_l)$ and $\mu_{ns} = \mu_l \lambda_l + \mu_g (1 - \lambda_l)$ are the no-slip density and viscosity correspondingly ($\rho_g$ and $\mu_g$ are gas density and viscosity).

Thus, in the chosen set of dimensionless parameters information about diameter is contained in the Reynolds and Froude numbers. Moreover, we note that in Beggs \& Brill and Mukherjee \& Brill correlations diameter number is also excluded from the models. 

Here we would like to justify the order of steps in the proposed method, where first we determine the volume fraction of the liquid, and only then we determine the flow regime. The logic is as follows. A typical flow regime map is built on the 2D plane in the axes being liquid and gas velocity numbers (or these parameters could be superficial velocities of gas and liquid) \cite{brill1999multiphase}, \cite{bratland2010pipe}. The velocity number is proportional to the superficial velocity which, in turn, by definition is the linear velocity times the volume fraction. From computational fluid dynamics, it is known that by solving the system of conservation laws with closure relations it is possible to find fields of gas and liquid linear velocities and liquid holdup. Hence, in order to determine where we are on the flow regime map, one needs to determine first the volume fraction of the fluid (i.e., holdup), and only then one determines the superficial velocities for this volume fraction, which allows determining the point on the flow regime map and, hence, to identify the flow regime. Therefore, in our method we first determine the holdup, then the flow regime, and not vice versa.

\subsubsection{Model for liquid holdup}
The surrogate model for estimation of the liquid holdup parameter in the pipe segment is a regression. The following set of input parameters is used: inclination angle of the segment, liquid and gas velocities numbers, viscosity number, Reynolds and Froude numbers, no-slip liquid holdup. To improve the predictive ability of a surrogate model, the input data is divided into three categories according to the inclination angle of the segment: data points are responsible for horizontal, uphill and downhill flows. For each group a separate surrogate model for liquid holdup prediction is constructed.


\subsubsection{Model for flow regime identification}
In this subsection, the second surrogate model will be discussed. It is represented by a multi-class classifier and predicts flow regime. Since in the lab dataset there are only four flow regimes distinguished (bubble, slug, annular mist and stratified regimes), the surrogate model determines this parameter among them. Flow pattern and value of liquid holdup are linked parameters; that is why the second one is included in input features of this classification model.
Among other input features inclination angle, dimensionless parameters (Eq.~\ref{eq2}), Reynold and Froude numbers, no-slip liquid holdup are included. Similar to the liquid holdup prediction model, in this case, three ML models for horizontal flow, upward and downward are constructed.
\subsubsection{Model for pressure gradient}
\label{Pressure gradient model}
In order to estimate pressure gradient in the segment the equation for conservation of linear momentum is used. From this conservation law, it is possible to express pressure gradient which consists of three components: friction, elevation and acceleration. Using the definition of friction component, it could be written in the form: 
$$
\left( \frac{dp}{dl}\right)_f = \frac{f \rho_{ns}v_m^2}{2d},$$

where $f$ is a friction factor. Further, elevation component is expressed in the form: 
$$\left( \frac{dp}{dl}\right)_{el} = \rho_{m}g \sin{\theta},
$$

where $\rho_{m} = \rho_l \alpha_l + \rho_g (1 - \alpha_l)$ is an in-situ density of a mixture (here $\alpha_l$ is a liquid holdup) and $\theta$ is an inclination angle. The final component is acceleration or kinetic energy that results from a change in velocity. In many cases this part of the pressure gradient is negligible, but it is significant in the case of compressible phase presence under relatively low pressures. Doing similar math to \cite{beggs1973study}, we express the acceleration component as $$\left( \frac{dp}{dl}\right)_{acc} = -\frac{\rho_m v_m v_{sg}}{p} \frac{dp}{dl},$$ where $p$ is pressure. Thus, the equation for pressure gradient has the following form:
\begin{equation}
\frac{dp}{dl} = -\frac{(\rho_l \alpha_l + \rho_g (1 - \alpha_l))g \sin{\theta} + \frac{f (\rho_l \lambda_l + \rho_g (1 - \lambda_l)) v^2_m}{2d}}{1 - \frac{(\rho_l \alpha_l + \rho_g (1 - \alpha_l)) v_m v_{sg}}{p}}.
\label{eq13}
\end{equation}

In Eq.~\ref{eq13} all parameters are known apart from the liquid holdup ($\alpha_l$) and friction factor ($f$) for a multiphase flow. Using the first surrogate model, the value of $\alpha_l$ could be defined. In order to estimate the friction factor, it is necessary to construct another regression model. Utilizing data from the laboratory database, we recalculate the model targeted values of friction factor by using Eq.~\ref{eq13}. In the input parameters of this model the following set is included: liquid velocity numbers, viscosity number, Reynolds and Froude numbers, liquid holdups ($\alpha_l$ and $\lambda_l$), relative roughness.  

At the end of this section, we would like to discuss an alternative approach, where the pressure gradient is directly predicted by the surrogate model trained on lab data. This approach is discussed in the earlier work~\cite{kanin2018method}, where authors represent the set of input parameters of this model, values of scores and cross-plot with results. Despite the results obtained are quite high, the entire method for pressure gradient prediction (liquid holdup, flow pattern and pressure gradient) provides very low scores on the field data, which was used for model validation. This model contains the diameter number among input parameters, which has narrow variation ranges for lab and field data that slightly overlap. Excluding this dimensionless diameter number, it is then not possible to predict the pressure gradient directly from the data. The other problem of directly prediсtiing the pressure gradient is its complex structure: gravitational, acceleration and friction components. Hence, in the present work we propose the new approach for pressure gradient estimation.

\subsection{Applied Machine Learning algorithms, their tuning and evaluation scores}

Let us introduce some notations. Matrix with input features is denoted as $\textbf{X}$ and has size $m\times d$ ($m$ is a number of samples and $d$ is a number of features), $y$ is a matrix (in the majority cases -- column) that contains target real values. It has size $m \times r$, where $r=1$ when we model a single output, and $r>1$ when we  consider a multiple output task. Moreover, $\hat{y}$ is a matrix with predicted values and has the same size as matrix $y$.

In the present paper four Machine Learning algorithms are applied: Random Forest, Gradient Boosting Decision Trees, Support Vector Machines and Artificial Neural Networks (this set of ML algorithms is the same as utilized in the article \cite{kanin2018method}). So, all these algorithms are used to construct each of the models described above for liquid holdup prediction, flow pattern identification and pressure gradient calculation in order to compare their predictive capability. Let us briefly describe these algorithms.

\textbf{Random Forest} \cite{breiman2001random} can be used to solve both regression and classification tasks by constructing ensembles of ML models: the algorithm constructs several independent decision trees and average them
\begin{equation}
h_N(\textbf{x}) = \frac{1}{N}\sum_{i=1}^{N}h_i(\textbf{x}),
\label{eq3}
\end{equation}
where $h_i(\textbf{X})$ is a decision tree, $N$ is a total number of decision trees. 

Each decision tree is trained on its own bootstrapped sample: we select randomly and  without replacement data points from the initial dataset to construct the bootstrapped sample. Moreover, as input features for this sample we select randomly a subset of the initial $d$ features.

During construction of surrogate models, the implementation of Random Forest algorithm in the Scikit-learn library \cite{scikit-learn} on Python language is utilized. For classification and regression problems we use functions \textit{RandomForestClassifier()} and \textit{RandomForestRegressor()} respectively.


All machine learning algorithms have their own set of parameters that should be tuned based on the utilized dataset. These parameters are called hyperparameters. In the case of the Random Forest algorithm, the following hyperparameters are adjusted: the number of trees in the forest (n\_estimators), maximum depth of the tree (max\_depth), number of features that algorithm considers in the process of tree construction (max\_features). The hyperparameters of the Random Forest algorithm are tuned in the same sequence as they are listed.

 \textbf{Gradient Boosting Decision Trees} \cite{friedman1999greedy}. This technique also constructs ensemble of ML models and can be used both for regression and classification. The algorithm constructs several decision trees and 
the final result is the weighted sum of them:
\begin{equation}
h_N(\textbf{x}) = \sum_{i=1}^{N}\alpha_i h_i(\textbf{x}).
\label{eq4}
\end{equation}
Each decision tree $h_i(\textbf{x})$ tries to fit anti-gradient of the loss function (logistic, exponential loss functions): 
$$-\frac{\partial L(f_{i-1}(\textbf{x}_j), y_j)}{\partial f_{i-1}(\textbf{x}_j)}\bigg|_{f_{i-1}(\textbf{x}_j) = \sum_{k=1}^{i-1}\alpha_kh_k(\textbf{\textbf{x}}_j)}, j = 1,..., m,$$ where $f_{i-1}(\textbf{x}_j) = \sum_{k=1}^{i-1}\alpha_kh_k(\textbf{\textbf{x}}_j)$ is a predicted value for a data point with index $j$ on the $(i-1)$-th boosting iteration.

Weights in the sum of decision trees ($\alpha_i$) are found by a simple line search:
$$\alpha_i = \textrm{argmin}_{\alpha>0} \sum_{j=1}^m L(f_{i-1}(\mathbf{x}_j) + \alpha h_i(\textbf{x}_j), y_j).$$

Besides Gradient Boosting algorithm, implemented in Scikit-learn library \cite{scikit-learn} in functions \textit{GradientBoostingClassifier()} (for classification models) and \textit{GradientBoostingRegressor()} (for regression models), XGBoost \cite{Chen:2016:XST:2939672.2939785} implementation (the latest one version) is also used. It is a standard gradient boosting with additional regularization to prevent over-fitting.

For the Gradient Boosting algorithm the following set of hyperparameters are tuned: the number of constructed trees (n\_estimators), learning rate that shrinks the contribution of each tree (learning\_rate), maximum depth of the tree (max\_depth), number of features that algorithm considers in the process of tree construction (max\_features), the fraction of samples that are used for learning each tree (subsample) and parameters concerning the building of the trees' structure (min\_samples\_split, min\_samples\_leaf). We tune the hyperparameters of the Gradient Boosting algorithm in the following sequence: n\_estimators, max\_depth, min\_samples\_split, min\_samples\_leaf, max\_features, subsample and learning\_rate.

The Gradient Boosting algorithm is often applied for solving different problems connected with oil and gas industry. For example, in paper \cite{makhotin2019gradient} authors utilized this method for prediction of the flow rate of the well after the process of hydraulic fracturing. In addition, in the article \cite{ignatov2018tree} authors created a model based on the Gradient Boosting algorithm for calculation of the bottomhole pressure in the case of transient multiphase flow in wells.

\textbf{Support Vector Machine} \cite{cortes1995support}. Similarly to the previous two algorithms, SVM could be used in classification and regression analysis. In case of SVM very often input \textbf{x} is transformed into a high-dimensional feature space by some non-linear mapping $F(\cdot)$. This transformation is applied in order to make data linearly separable (in case of classification) or to make it possible to fit transformed data with a linear function  $\textbf{w} \cdot F(\textbf{x}) + b$, which is nonlinear in the original input space.

It can be proved that instead of explicitly defining $F(\cdot)$ it is sufficient to define a kernel  $K(\textbf{x}, \textbf{x}')$, being a dot product  between $\textbf{x}$ and $\textbf{x}'$ in the new input space, defined by the transformation $F(\cdot)$:
$$K(\textbf{x}, \textbf{x}') = F(\textbf{x}) \cdot F(\textbf{x}').$$

When constructing SVM we use a Gaussian kernel with width $\sigma$:
$$K(\textbf{x}', \textbf{x}) = \exp\left(\frac{||\textbf{x} - \textbf{x}'||^2}{2\sigma^2}\right).$$

SVM algorithm maximizes a classification margin, equal to $1 / \|\textbf{w}\|$, i.e. minimizes the norm $\|\textbf{w}\|$. In  case of binary classification, the algorithm also minimizes the sum of slack variables that are responsible for misclassification. The optimization problem has the following form:
\begin{align}
\min_{\textbf{w}, b, \xi} & \frac{||\textbf{w}||^2}{2} + C\sum_{i = 1}^{m} \xi_i \nonumber \\
\textrm{subject to:}\, & \: y_i (\textbf {w} F(\textbf x_i) + b) \geq 1 - \xi_i \: \textrm{and} \: \xi_i \geq 0, \: i\in [1, m].
\label{eq5}
\end{align}

In case of regression, the algorithm minimizes the sum of slack variables that are responsible for data points lying outside the $\epsilon-$tube and minimizes complexity term $\|\mathbf{w}\|^2$. The formulation of this optimization problem is as follows:
\begin{align}
\min_{\textbf{w}, b, \xi} & \frac{||\textbf{w}||^2}{2} + C\sum_{i = 1}^{m} (\xi_i + \xi_i^*) \nonumber \\ 
\textrm{subject to:}\,& \: y_i - \textbf{w} F(\textbf x_i) - b \leq \epsilon + \xi_i, \nonumber \\
&\textbf{w} F(\textbf x_i) + b - y_i\leq \epsilon + \xi_i^*, \nonumber \\ 
&\xi_i, \xi_i^* \geq 0, \: i \in [1, m].
\label{eq6}
\end{align}

Further, these optimization problems (\ref{eq5}, \ref{eq6}) are solved in dual formulation with the use of Lagrangian.

The SVM algorithm is also implemented in the Scikit-learn \cite{scikit-learn} in functions \textit{SVC()} and \textit{SVR()} for regression and classification tasks respectively, and these functions are used in the present work.

In the case of the SVM algorithm, the width of the Gaussian kernel ($\sigma$) and regularization parameter (C) are adjusted in the process of models construction. We tune these parameter together meaning that different combinations of $\sigma$ and C are considered.

\begin{figure}[!htb]
	\centering
	\includegraphics[width = 0.45\textwidth]{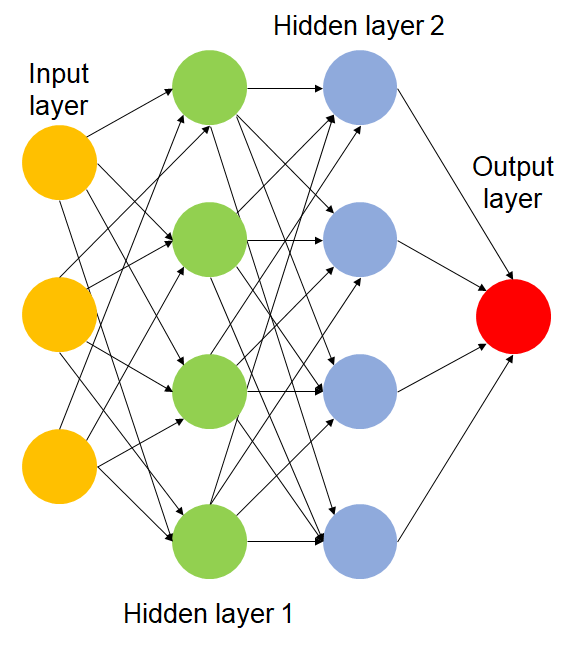}
	\caption{The schematic picture of artificial neural network}\label{fig27}
\end{figure}

\textbf{Artificial Neural Networks} \cite{rosenblatt1958perceptron}. The schematic picture of ANN is represented in Fig.~\ref{fig27}.
This ML algorithm could be applied in classification and regression problems. ANN consists of several layers (input, output and hidden). Each layer has a certain number of nodes. The first layer contains input parameters, so the number of nodes is $p$. The last one contains output parameters: in the case of regression problems or binary classification there is only one node. In the case of multiclass classification, the number of nodes is equal to the number of classes. The algorithm consists of forward and backward propagation. In the forward propagation process, the algorithm calculates values at each node in the following way: in order to obtain the value at the node $k$ in the layer $h+1$ the algorithm carries out a linear combination of values in nodes in layer $h$ with some weights. Then it applies to this linear combination activation function $g(\cdot)$:
\begin{equation}
\textbf{y}_{k, h+1} = g(\theta_{k, h \rightarrow h+1} ^ T \textbf{y}_h).
\label{eq7}
\end{equation}
In the backward process, the algorithm adjusts the weights by using a gradient descent optimization algorithm in order to decrease the value of loss function $L(f_i, y_i)$.

In classification problems sigmoid $\left(g(x) = \frac{1}{\exp(-x) + 1}\right)$ or hyperbolic tangent ($g(x) = \tanh(x)$) activation functions are used. In case of regression ReLu $(g(x) = \max(0, x))$ is often applied. Approaches to  initialize parameters of ANN model are listed in \cite{ANNInit2016}. In \cite{HDA2013} they discuss efficient ANN training algorithms, and in \cite{Ensembles2013} approaches to construct  ensembles of ANNs.

Artificial Neural Network is implemented in many libraries on Python language, for example, Keras \cite{chollet2015keras}, Theano \cite{2016arXiv160502688short}, Pyrorch \cite{paszke2017automatic}. In this paper the functions from Scikit-learn library \cite{scikit-learn} are used: for classification model: \textit{MLPClassifier()} and for regression model: \textit{MLPRegressor()}. Keras, Theano and Pytorch libraries are mainly oriented for the realization of convolution neural networks. That is why their usage is not necessary in the present simple case. 

Similar to the Gradient Boosting algorithm, the Artificial Neural Network is also relatively popular for solving oil and gas industrial problems. For example, it was used in~\cite{spesivtsev2018predictive} to predict the value of bottomhole pressure in a transient multiphase well flow. In the other paper~\cite{erofeev2019}, the authors applied different ML algorithms for predicting porosity and permeability alteration due to scale depositing, and it was shown that ANN has the best predictive ability for this problem. Some other studies with ANN models have been already mentioned in the Section \ref{Introduction}.

When applying neural networks to construct surrogate models we tune the following  hyperparameters: the number of hidden layers and number of nodes in hidden layers (hidden\_layer\_sizes), the value of learning rate (learning\_rate\_init). Also, activation function (activation) and a method of gradient descent (solver) are adjusted. Let us consider the approach of tuning the number of nodes in the hidden layers. Firstly, we take the ANN with one hidden layer and tune the number of nodes ($n_1$) in it. Further, we consider the ANN with two hidden layers and tune the number of nodes in the second layer ($n_2$) presuming that the number of nodes in the first layer equals $n_1$. The same procedure is carried out for the third layer. We also always check that adding of the new layer improves the value of metric; otherwise the new layer is not inserted. The maximum number of hidden layers in our models equals to three. There is no sense to utilize deeper ANN due to relatively small data set. After tuning the network structure, we tune the learning\_rate\_init parameter. Finally, we note that for the gradient descent in the back propagation procedure we utilize the Adam optimizer \cite{kingma2014adam}.

For tuning hyperparameters of ML algorithms, we use the grid search algorithm and the $M$ cross-validation procedure. Let us start with the description of the grid search. First of all, it is necessary to choose the sets of the possible values of hyperparameters. The grid search procedure passes through all various combinations of values of hyperparameters and on each iteration it trains the ML algorithm on the training dataset and tests the trained algorithm on the testing part on the dataset. Finally, the grid search algorithm chooses the optimal combination of hyperparameters that yields the best score on the testing dataset. Under the term score, we mean the value of the metric that is applied in the considered regression or classification problem. The set of utilized metrics in our work will be described further. 

Instead of dividing the data set on the training and the testing part just once, we apply cross-validation technique. In this procedure the dataset is divided into $M$ equal parts and $M - 1$ partitions are utilized as a training dataset and the remaining partition as a testing (it could be also named as validation) dataset. This process is repeated $M$ times, and on each iteration, different test partition is used. As a result, it is possible to compute $M$ test scores and average them. Further, the averaged score is utilized in the grid search algorithms for identifying what set of hyperparameters is the optimal.

In order to evaluate the score of the tuned model with the best hyperparameters the $N \times M$ cross-validation procedure is applied. In this method, a dataset is divided into $M$ equal parts $N$ times and at each $N$-step the partitions are different. Thanks to this technique calculated score of the model ($\mu$ --- mean) is more accurate as compared to the $M$ cross-validation and it also allows constructing confidence intervals of this score $\left(\pm \: \frac{1.96 \cdot \sigma}{\sqrt{M \cdot N}}\right)$ where $\sigma$ is a standard deviation. In this paper in case of the $M$ cross-validation $M = 5$ is used (this value is the most common), and in case of $N \times M$ cross-validation we set $M = 5$ and $N = 20$.   

The following set of evaluation metrics is used. In case of multi-class classification:
\begin{itemize}
\item F1-macro metrics
\begin{align}
P &= \frac{1}{|Y|}\sum_y \frac{TP_y}{TP_y + FP_y}, R = \frac{1}{|Y|}\sum_y \frac{TP_y}{TP_y + FN_y},\notag\\
F_1 & = \frac{2PR}{P+R}, 
\label{eq8}
\end{align}
where $P$ is a precision, $R$ is a recall, $|Y|$ is a number of classes, $TP_y$ --- true positives of class $y$, $FP_y$ --- false positives of class $y$ and $FN_y$ --- false negatives of class $y$.
\item Accuracy:
\begin{equation}
\textrm{Accuracy} = \frac{1}{m} \sum_{i=1}^m 1_{f(x_i) = y_i} = \sum_y \frac{TP_y + TN_y}{TP_y + FN_y + TP_y + FN_y},
\label{eq9}
\end{equation}
where $TN_y$ -- true negatives of class y.
\end{itemize}

In case of regression:
\begin{itemize}
\item Coefficient of determination ($R^2$-score)
\begin{equation}
R^2 = 1 - \frac{\sum_{i=1}^m (y_i - \hat{y}_i)^2}{\sum_{i=1}^m (y_i - \bar{y})^2},
\label{eq10}
\end{equation}
where $\bar{y} = \frac{1}{m}\sum_{i=1}^m y_i$.
\item Mean squared error
\begin{equation}
\textrm{MSE} = \frac{1}{m}\sum_{i=1}^m (y_i - \hat{y}_i)^2.
\label{eq11}
\end{equation}
\item Mean absolute error
\begin{equation}
\textrm{MAE} = \frac{1}{m}\sum_{i=1}^m |y_i - \hat{y}_i|.
\label{eq12}
\end{equation}
\end{itemize}

\section{Lab data base: sources and structure. Preprocessing}
\label{lab data set}

For construction of surrogate models, namely, for training, validation and testing stages, the lab data set is collected from open source articles, PhD theses and books. Some part of this dataset has been used in the earlier study~\cite{kanin2018method}.

From paper \cite{minami1987liquid} 111 data point for horizontal flow are taken. The author of this paper carried out experiments using kerosene and water as a liquid phase and air as a gas phase. The next 88 data points are from the article \cite{abdul1996liquid} in which the author performed the experiment for horizontal flow using kerosene and air. Further, 1400 dataset are taken from \cite{mukherjee1979phd} which consists of uphill, downhill and horizontal flows in pipes. So in this case the inclination angle varies from $-90^\circ$ to $90^\circ$. The author used kerosene and lube oil as a liquid phase and air as a gas phase in the experiments. From \cite{eaton1966phd} 238 data points of horizontal multiphase flow of water and natural gas are used in the collected dataset. The next 188 data points of water and air multiphase flow are taken from \cite{beggs1973phd}. The data includes not only horizontal flows but also pipes at angles from $-10^{\circ}$ to $10^{\circ}$. And the final 535 data point are taken from \cite{AndritsosPhD} in which the author conducted experiments with air-water and air-glycer multiphase horizontal flow.   

As a result, the consolidated data base contains the total of 2560 points for constructing the ML model for liquid holdup prediction and flow pattern identification. Out of the total number, approximately 1700 points contain information about the value of the measured pressure gradient. In the process of constructing the surrogate model for pressure gradient, we recalculate the values of the friction factor according to Eq.~\ref{eq13} as mentioned in the subsection \ref{Pressure gradient model}. For some data samples we obtained non-physical values (e.g., less than zero or too large as compared with Fanning friction factor for mixture). That is why we do not use these data points when constructing surrogate models. As a result, the total number of data points for training, validation and testing of the resultant surrogate model is approximately 1300. In the collected dataset the flow pattern is provided for about 1400 data points. In order to fill the remaining samples the flow pattern map created by Mukherjee \cite{mukherjee1983liquid} is used.

\begin{figure}[h!]
	\centering
	\includegraphics[width=0.49 \textwidth]{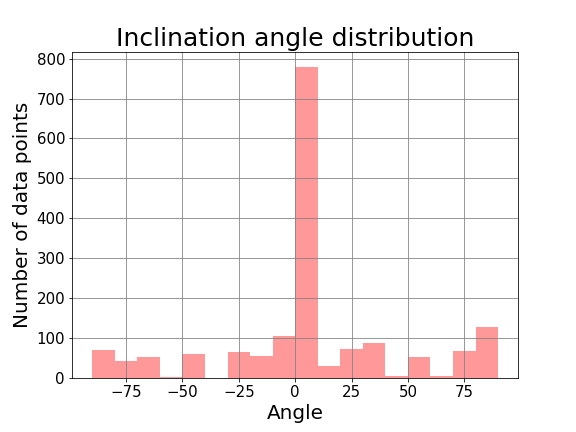}
	\caption{Histogram with distribution of inclination angles}\label{fig5}
\end{figure}

As a result of the exploratory analysis of the dataset we provide the following charts. 
First of all, let us observe the distribution of angles at which the pipes are orientated in the experiments. In Fig.~\ref{fig5} the distribution of inclination angles is plotted.

From this diagram, one could find that the majority of experiments in this dataset were conducted with horizontal pipes. The remaining part consists of experiments in which pipes are oriented at angles: $0^{\circ} < |\theta| \leq 90^{\circ}$. The number of measurements in uphill flows is equal to ones in downhill flows due to the construction of laboratory equipment.

\begin{figure}[h!]
	\centering
	\includegraphics[width=0.49 \textwidth]{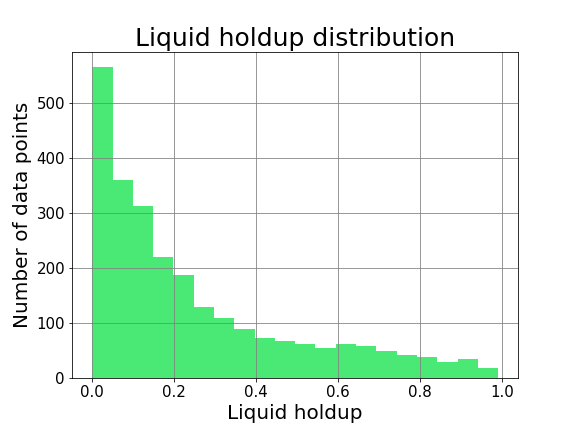}
	\caption{Histogram with distribution of liquid holdup}\label{fig3}
\end{figure}

Further, the distribution of values of liquid holdup is considered. This distribution is illustrated in Fig.~\ref{fig3}.

From this chart, one could clarify that the mean value of the liquid holdup parameter in the dataset is equal to 0.25. This value indicates that in the majority of the experiments researches used the larger amount of air compared with the volume of liquid.   

\begin{figure}[h!]
	\centering
	\includegraphics[width=0.49 \textwidth]{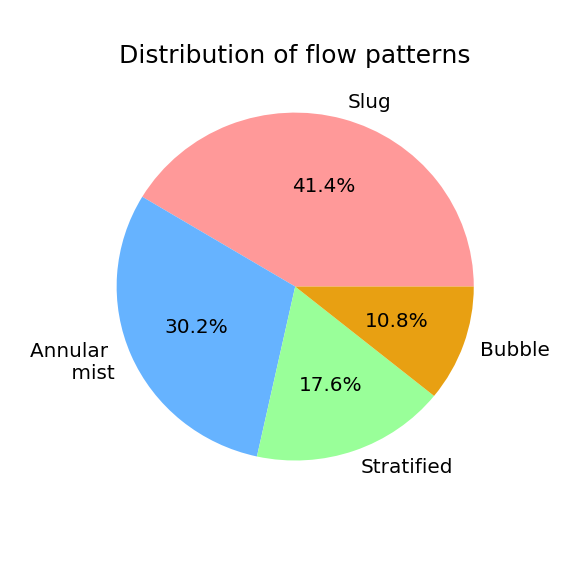}
	\caption{Pie chart with flow pattern distribution}\label{fig2}
\end{figure}

Let us move on to the distribution of flow patterns in the dataset. In Fig.~\ref{fig2} the pie chart with this distribution is represented. 

About 41 \% of the data points belong to the slug flow regime. Approximately 30 \% of the data is annular mist flow type, slightly larger than 17 \% of the samples are stratified flow pattern. The minority class is bubble flow. These numbers indicate that the dataset is imbalanced and, as a result, in case of multi-class classification data points of slug and annular mist flow regimes could influence significantly the training process leading to the miss-classification errors. In order to balance data and improve classification accuracy approaches to imbalanced classification can be used \cite{Imbalance2015,Imbalance2019}.

\begin{figure}[h!]
	\centering
	\includegraphics[width=0.49 \textwidth]{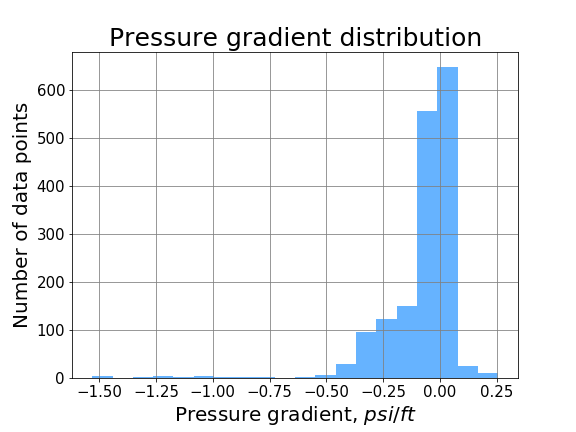}
	\caption{Histogram with pressure gradient distribution}\label{fig4}
\end{figure}

Finally, the distribution of pressure gradients is examined. The diagram with this parameter is represented in Fig.~\ref{fig4}. 

The mean value of pressure gradient loss in the dataset equals to -0.088 psi/ft. The maximum value of the pressure gradient is 0.25 psi/ft that occurred in the case of the orientation of the segment at $\theta = -90^{\circ}$. The minimum value is observed in the horizontal case ($\theta = 0^{\circ}$) and equals to $-1.53$ psi/ft.

At the end of the discussion about lab data set, the Table~\ref{tab1} with parameters and their ranges from the collected data set is represented.
\begin{table*}[h!]
\begin{center}
\begin{tabular}{|C{3.5cm}|C{2cm}|C{2cm}|C{2cm}|C{2cm}|C{2cm}|C{2cm}|}
\hline
Researcher & No. points & Angle & $v_{sg}$, ft/s & $v_{sl}$, ft/s & P, psi& T, F \\
\hline
Minami and Brill \cite{minami1987liquid}& 111 & 0& 1.56 -- 54.43 & 0.02 -- 3.12 & 43.7 -- 96.7 & 76 -- 118\\

Abdul \cite{abdul1996liquid} & 88 &  0 & 0.64 -- 160.46 & 0.01 -- 5.99 & 28.6 -- 133.3 & 82 -- 120 \\

Beggs \cite{beggs1973phd} & 188 & -10 -- 10 & 0.98 -- 83.1 & 0.07 -- 17.07 & 35 -- 98.9 & 38 -- 90 \\

Eaton \cite{eaton1966phd} & 238 & 0 & 0.37 -- 71.85 & 0.04 -- 6.92 & 290.6 -- 854 & 57 -- 112\\

Mukherjee \cite{mukherjee1979phd} & 1400 & -90 -- 90 & 0.12 -- 135.53 & 0.05 -- 14.31 & 20.9 -- 93.7 & 18 -- 165.5\\
Andritsos \cite{AndritsosPhD} & 535 & 0 & 2.62 - 535.56 & 0.001 - 1.09 & 14.25 - 28.44 & 50 - 79.7\\
\hline
All dataset & 2560 & -90 -- 90 & 0.12 -- 535.56 & 0.001 -- 17.07 &20.9 -- 854 & 18 -- 165.5\\
\hline
\end{tabular}
\end{center}
\end{table*}
\begin{table*}[h!]
\begin{center}
\begin{tabular}{|C{3.5cm}|C{2cm}|C{2cm}|C{2cm}|C{2cm}|C{2cm}|C{2.5cm}|}
\hline
Researcher & d, ft &$\rho_l$, $\mathrm{lb/ft^3}$ & $\mu_l$, cP & $\sigma_{lg}$, Dynes/cm & $\rho_g$, $\mathrm{lb/ft^3}$ & $\mu_g$, cP\\
\hline
Minami and Brill \cite{minami1987liquid} & 0.25 & 49.4 -- 62.4 & 0.58 -- 2 & 26.01 -- 72.1 & 0.212 -- 0.484 & 0.0096 -- 0.0104  \\
Abdul \cite{abdul1996liquid} & 0.167 & 49.3 -- 50.2 & 1.29 -- 1.98 & 23.5 -- 26.3 & 0.134 -- 0.632 & 0.0097 -- 0.0105 \\
Beggs \cite{beggs1973phd} & 0.083, 0.125& 62.4 & 0.78 -- 1.4 & 70.8 -- 75.6 & 0.181 -- 0.512 & 0.0173 -- 0.0187 \\
Eaton \cite{eaton1966phd} & 0.17 & 62.9 -- 63.6 & 0.71 -- 1.33 & 61.6 -- 66.5 & 0.877 -- 2.866 & 0.0111 -- 0.0127 \\
Mukherjee \cite{mukherjee1979phd} & 0.125 & 48.1 -- 54.1 & 0.63 -- 74.4 & 20.9 -- 37.5 & 0.101 -- 0.47 & 0.0085 -- 0.082\\
Andritsos \cite{AndritsosPhD} & 0.08, 0.31 & 62.43 - 76.16 & 1 - 80 & 66 - 73 & 0.072 -- 0.147 & 0.0177 -- 0.0185\\
\hline
All dataset & 0.08 -- 0.31& 48.1 -- 63.6 & 0.58 -- 80 & 20.9 -- 75.6 & 0.072 -- 2.866 & 0.0085 -- 0.082  \\
\hline
\end{tabular}
\end{center}
\end{table*}
\begin{table*}[h!]
\begin{center}
\begin{tabular}{|C{3.5cm}|C{2cm}|C{6cm}|C{2cm}|}
\hline
Researcher & Liquid holdup & Flow pattern & $\frac{dP}{dL}$, psi / ft\\
\hline
Minami and Brill \cite{minami1987liquid} & 0.008 -- 0.45 & Slug, Annular mist, Stratified & - \\
Abdul \cite{abdul1996liquid} & 0.009 - 0.61 & Slug, Annular mist, Stratified & -\\
Beggs \cite{beggs1973phd} & 0.02 -- 0.88 & Slug, Bubble, Annular mist, Stratified & -0.38 -- 0.027\\
Eaton \cite{eaton1966phd} & 0.006 -- 0.73 & Slug, Bubble, Annular mist, Stratified & -\\
Mukherjee \cite{mukherjee1979phd} & 0.01 -- 0.99 & Slug, Bubble, Annular mist, Stratified & -0.47 -- 0.25\\
Andritsos \cite{AndritsosPhD} & 0 - 0.67 & Slug, Bubble, Annular mist, Stratified & -1.53 - 0\\
\hline
All dataset & 0 - 0.99 & Slug, Bubble, Annular mist, Stratified & -1.53 -- 0.25\\
\hline
\end{tabular}
\caption{Parameters and their ranges from collected data set}
\label{tab1}
\end{center}
\end{table*}

Now, we turn to the data preprocessing. After collecting the lab data from aforesaid different sources, we combine all parts. Further, dimensionless parameters from section \ref{input parameters} required for constructing surrogate models are calculated. When selecting the data for training Machine Learning algorithms, it is necessary to analyze possible outliers. These data points are isolated from others in the space of input features. That is why ML algorithm could not predict these point normally. Moreover, outliers could influence on ML algorithms' training process. For identification of possible outliers we used the following techniques. 

The first one is construction of boxplots for each feature. Using each boxplot, one could identify data points that lie far from the main part of the sample according to values of the considered feature. However, these separated points could be located quite close to each other and, as a result, ML algorithm predicts them without significant error. That is why, using boxplots, it is feasible to suggest potential outliers.      

The other approach for outlier detection is Machine learning algorithms from scikit-learn library such as $LocalOutlierFactor()$ and $IsolationForest()$. 
These methods are targeted for finding isolated points with the help of k-nearest neighbours and forest algorithms correspondingly. However, it is necessary to tune hyperparameters of these algorithms. One of the possible ways to perform tuning is usage result of the surrogate model trained on the whole dataset. Knowing data points with the large errors of prediction, we tune hyperparameters and after that compare results of $LocalOutlierFactor()$, and $IsolationForest()$ with poor predicted data points. To further automate hyperparameters tuning and model selection for anomaly detection we can use approaches from \cite{ModelSelection2015}.

For validation of outlier detection, these data points are deleted from the dataset, another surrogate model is constructed, and the model's score is compared with the previous result.

\section{Results and Discussion}
Here we present histograms and tables with methods scores, cross-plots (for regression problems) and confusion matrix (for classification problem). Also, models with the best predictive capability are highlighted.

\begin{figure}[h!]
	\centering
	\includegraphics[width=0.5 \textwidth]{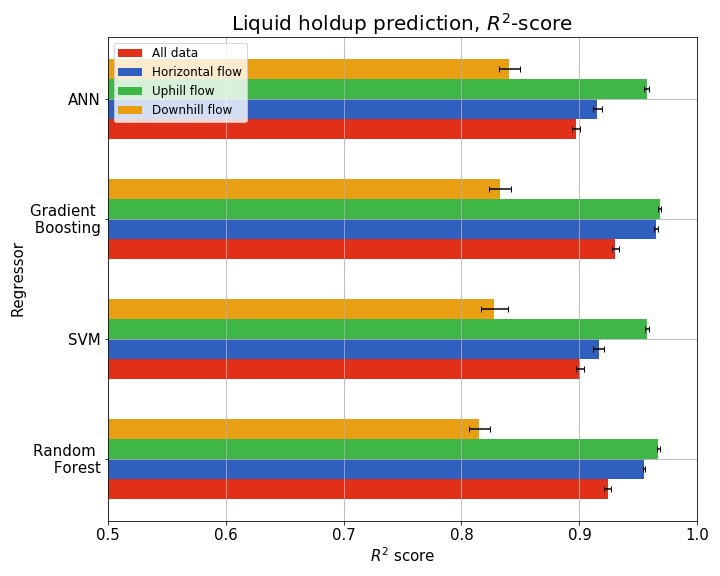}
	\caption{Histogram with $R^2$-scores in the model for liquid holdup prediction}\label{fig6}
\end{figure}

First of all, the results of the model for liquid holdup prediction will be discussed. Histogram with $R^2$ scores that investigated ML algorithms provide is depicted in Fig.~\ref{fig6}. 

In this histogram, one could observe not only the results of scores of models for horizontal, uphill and downhill flows but also scores for models that are trained using all dataset.

From this bar chart, one could find that the Gradient Boosting has the best predictive capability in all cases apart from the downhill flow (the version of XGBoost and Gradient Boosting function from Scikit-learn library provides approximately the same results). When all data set is used for training, validation and testing model, Gradient Boosting has $R^2 = 0.931 \pm 0.003$. In the model for horizontal flow $R^2 = 0.965 \pm 0.002$, for uphill flow $R^2 = 0.968 \pm 0.011$. In the case of downhill flow, the better result is obtained by using ANN with $R^2 = 0.841 \pm 0.009$. From these results, the following conclusion could be made: models for horizontal and uphill flows have a very high coefficient of determination, while the model for downhill flow demonstrates a low $R^2$ score. 

\begin{figure}[h!]
	\centering
	\includegraphics[width=0.5 \textwidth]{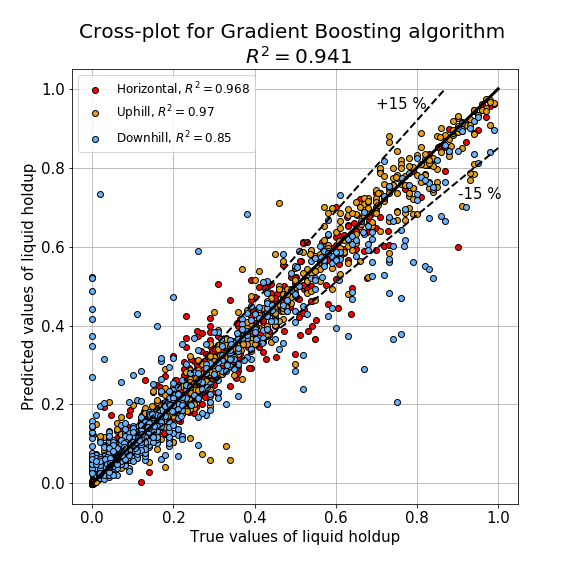}
	\caption{Cross-plot for the Gradient Boosting algorithm in the liquid holdup model. Different colours are responsible for different categories by inclination angles}\label{fig7_1}
\end{figure}

\begin{figure}[h!]
	\centering
	\includegraphics[width=0.5 \textwidth]{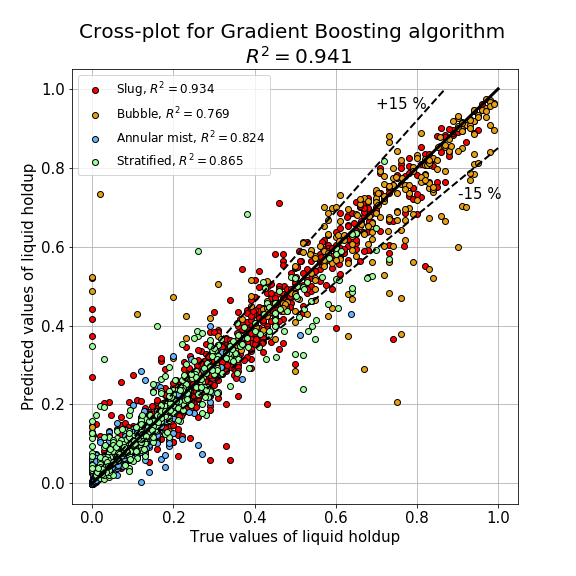}
	\caption{Cross-plot for the Gradient Boosting algorithm in the liquid holdup model. Different colours are responsible for different flow regimes (observed in the experiment)}\label{fig7_2}
\end{figure}

\begin{figure}[h!]
	\centering
	\includegraphics[width=0.5 \textwidth]{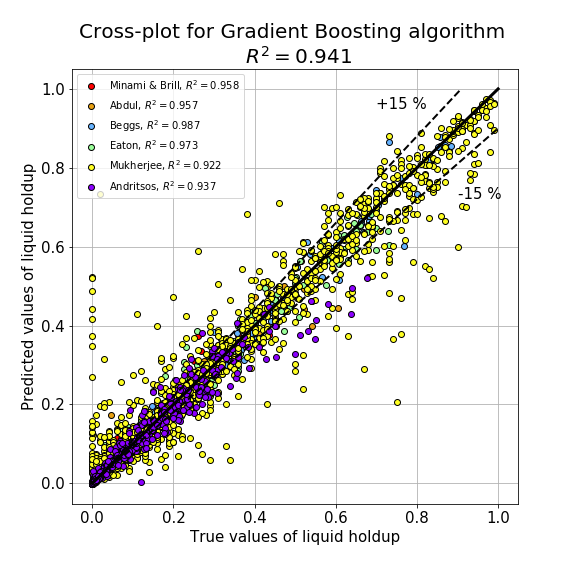}
	\caption{Cross-plot for the Gradient Boosting algorithm in the liquid holdup model. Different colours are responsible for different experimentators}\label{fig7_3}
\end{figure}

Since the liquid holdup model is a regression one, it is also important to construct a cross-plot in order to understand the number of possible outliers. In this graph on the X-axis real values are plotted while predicted values are on the Y-axis. In Fig.~\ref{fig7_1} - \ref{fig7_3} such cross-plots are demonstrated. 

In order to construct these plots, we predict liquid holdup values using surrogate models for horizontal, uphill and downhill flows based on Gradient Boosting algorithm. For this prediction $cross \_ val \_ predict (\cdot)$ function from scikit-learn Python library is utilized. Further, predicted values are combined into one table, and the dataset with predicted and initial values of liquid holdup is divided into categories with respect to different features, for example, flow patterns, experimentators and flow directions. Each category is marked by different colours on cross-plots (Fig.~\ref{fig7_1} - \ref{fig7_3}) and also in each category the coefficient of determination is found.

By analyzing these plots, one could observe that certain data samples are poorly predicted (with error more than 15\%). Mainly these points belong to downhill flow represented in Mukherjee experimental data. In terms of flow patterns, the worst predicted points belong to bubble and annular mist regimes. Besides Mukherjee data set, the large part of predictions for Andritsos data have errors exceeding 15\% boundary.

In order to analyze outliers, we apply Machine Learning techniques mentioned at the end of Section \ref{lab data set} and also construct boxplots. Using these methods, one could find out that a relatively small number of supposed isolated data points have large errors of predictions obtained by Gradient Boosting algorithms. As a result, there are two possible reasons for large errors: ML algorithm could not find out required dependence for these points or values of the liquid holdup in these cases were measured with instrumental error.  


All tree-based ML algorithms allow obtaining feature importance. It is a characteristic that shows the significance of each feature in the prediction of the output value. The rank of each feature is calculated according to the following idea: importance is computed for a single decision tree by the sum of weighted values that define how considered attribute improves the information criterion (Gini index, Information gain or others) when it is used in the splitting. Each weight in this sum is a number of samples in the divided node. After that ranks are averaged across all constructed decision trees and the final ranks are obtained. 

So, the Gradient Boosting algorithm shows the following sequence of feature importance in the model for liquid holdup prediction constructed using the whole  dataset (from the most to the least influential):
\begin{enumerate}
\item No-slip liquid holdup
\item Angle of inclination
\item Gas velocity number
\item Liquid velocity number
\item Reynolds number
\item Froude number
\item Viscosity number
\end{enumerate}

To compare predictive capability of the created model for liquid holdup calculation with existing methods we choose among 
empirical correlations Beggs \& Brill and Mukherjee \& Brill. Among mechanistic models we use TUFFP Unified, the combination of Ansari (when the module of inclination angle greater than $45^{\circ}$) and Xiao models (when $|\theta| < 45^{\circ}$). In addition, we also utilize Leda Flow point model and OLGAS. In Table~\ref{tab3} we represent cases where these methods could be applied. The following coefficients of determination are obtained: Beggs \& Brill (0.81), Mukherjee \& Brill (0.68), TUFFP Unified (0.834), combination of Ansari and Xiao (0.841), Leda (0.88), OLGAS (0.891). Using these scores, it is possible to identify that Leda Flow and OLGAS models yield the most accurate results. Further, we construct cross-plots with results of calculations via these models.
\begin{table*}[h]
\begin{center}
\begin{tabular}{ | c | c | c | c |}
\hline
Correlation / Mechanistic model &  Inclination angle & Diameter\\ 
\hline
Beggs \& Brill (1973) & from $-90^{\circ}$ to $90^{\circ}$ & arbitrary\\ 
\hline
Mukherjee \& Brill (1985) & from $-90^{\circ}$ to $90^{\circ}$ & arbitrary\\ 
\hline
Ansari (1994) & vertical / predominantly vertical & suitable for wells\\ 
\hline
Xiao (1994)  & horizontal / near horizontal & suitable for pipelines\\ 
\hline
TUFFP Unified (2003) & from $-90^{\circ}$ to $90^{\circ}$ & arbitrary\\ 
\hline
OLGAS (2002) & from $-90^{\circ}$ to $90^{\circ}$ & arbitrary\\ 
\hline
Leda Flow point model (2005) & from $-90^{\circ}$ to $90^{\circ}$ & arbitrary \\ 
\hline
\end{tabular}
\caption{Empirical correlations and mechanistic models best practices. All the mentioned methods could be applied for calculation of parameters for multiphase flow of oil, gas or gas condensate.}
\label{tab3}
\end{center}
\end{table*}

\begin{figure}[h!]
	\centering
	\includegraphics[width=0.5 \textwidth]{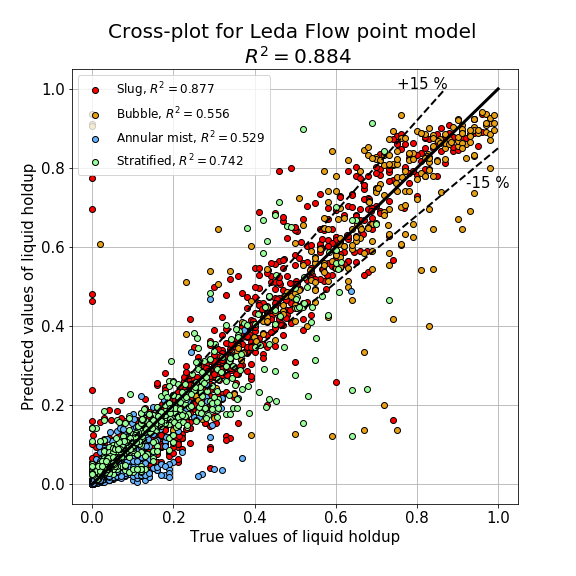}
	\caption{Cross-plot for results of Leda Flow point model. In this plot, different flow regimes are marked by different colours}\label{fig15}
\end{figure}

In Fig.~\ref{fig15} the cross-plot with result estimated by Leda Flow point model is represented. This mechanistic model yields the coefficient of determination equal to 0.884 on the collected lab data set that is less accurate than Gradient Boosting algorithm provides in this case. From Fig.~\ref{fig15} it is possible to identify that the worst predicted data points by Leda Flow model belong to bubble and annular mist flow patterns.

\begin{figure}[h!]
	\centering
	\includegraphics[width=0.5 \textwidth]{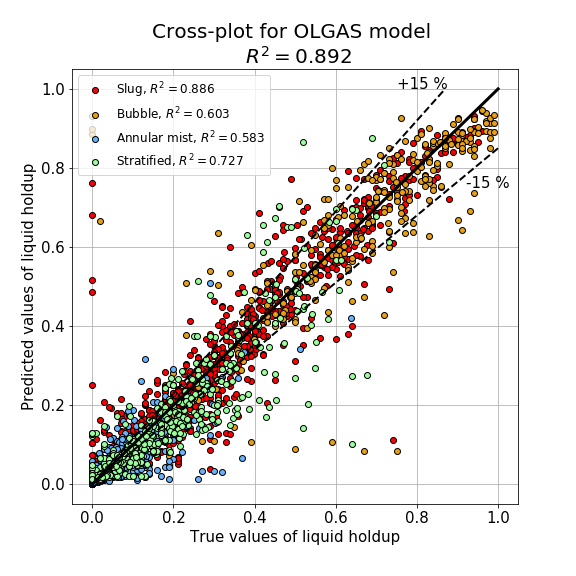}
	\caption{Cross-plot for results of OLGAS model. In this plot, different flow regimes are marked by different colours}\label{fig15_1}
\end{figure}

The best coefficient of determination on the lab dataset is obtained by OLGAS model. In Fig.~\ref{fig15_1} the cross-plot is depicted. From this graph, one could find out that the $R^2$ in the case of usage OLGAS model is equal to 0.89 that is also less accurate than the value of coefficient of determination in the model constructed with the help of the Gradient Boosting algorithm. Similarly to Leda Flow model, the worst predicted data points by OLGAS model belong to bubble and annular mist flow patterns.    

\begin{figure}[h!]
	\centering
	\includegraphics[width=0.5 \textwidth]{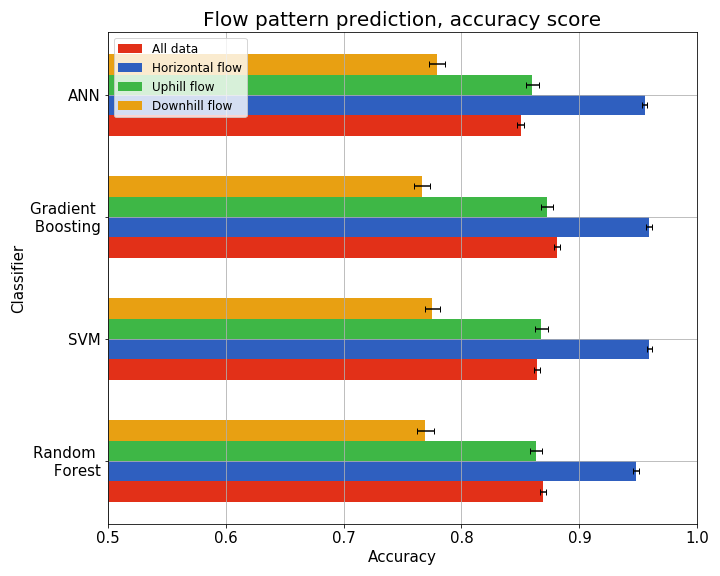}
	\caption{Histogram with accuracy scores in the model for flow pattern prediction}\label{fig8}
\end{figure}

Further, let us move on to the results of the second model that predicts flow patterns in the segment. Similarly to the discussion of the liquid holdup model, we begin with the histogram with accuracy scores for all ML algorithms. This bar chart is plotted in Fig.~\ref{fig8}. 

The Gradient Boosting algorithm has the best predictive capability when all data set is used for model creation with an accuracy score $0.88 \pm 0.003$. In the models for horizontal flow and uphill flow also Gradient Boosting has the best accuracy scores that equal to $0.959 \pm 0.003$ and $0.872 \pm 0.005$, correspondingly. Finally, in the case of downhill flow ANN demonstrates the best predictive capability with an accuracy score of $0.78 \pm 0.007$. From these results, one could conclude that the models for horizontal and uphill flows have good accuracy scores, while the model for downhill flow generates poor results.

Using predicted classes by surrogate models for horizontal, upward and downward flows, it is possible to categorize these results according to a data source and find accuracy score for each experimentator: Minami \& Brill (0.937), Abdul (0.989), Beggs (0.936), Eaton (0.987),  Mukherjee (0.828), Andritsos (0.981). From these values one could identify that predictions for Mukherjee dataset contain the largest number of misclassified data points because this dataset contains experiments for downhill flow which have poor predictions compared with horizontal and uphill flows (Fig.~\ref{fig8}). 

\begin{figure}[h!]
	\centering
	\includegraphics[width=0.5 \textwidth]{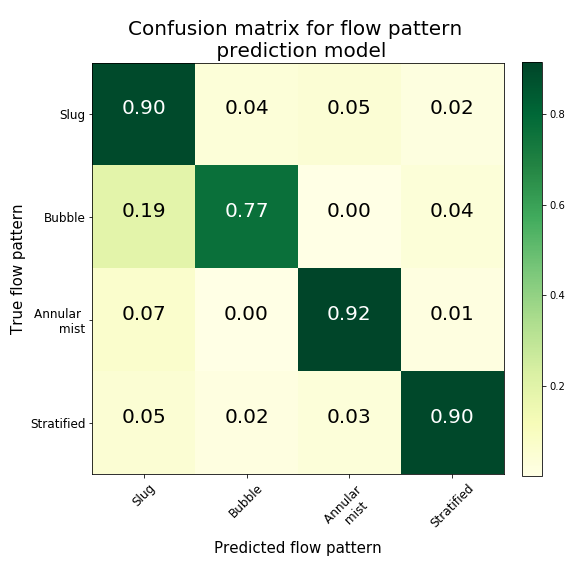}
	\caption{Confusion matrix for results of flow pattern prediction model}\label{fig9}
\end{figure}

For classification problem, it is possible to construct a confusion matrix. Each row of this matrix represents samples that belong to true class labels while each column represents predicted class labels. Using the confusion matrix, one could find how well ML algorithm predicts each class. In the Fig.~\ref{fig9} the confusion matrix is depicted. The values in this matrix relate to the result of flow pattern prediction model in which the Gradient boosting algorithm and all dataset are used.

Analyzing the confusion matrix in Fig.~\ref{fig9}, it is possible to reveal that the Gradient Boosting algorithm predicts quite well annular mist, slug and stratified flow patterns and the worst defines bubble flow regime.

The list of feature importance in the model for flow pattern prediction is the following:
\begin{enumerate}
\item No-slip liquid holdup
\item Gas velocity number
\item Angle of inclination
\item Froude number
\item Liquid velocity number
\item Viscosity number
\item Liquid holdup
\item Reynolds number
\end{enumerate}
This model is based on the Gradient Boosting algorithm constructed using the whole data set.

\begin{figure}[h!]
	\centering
	\includegraphics[width=0.5 \textwidth]{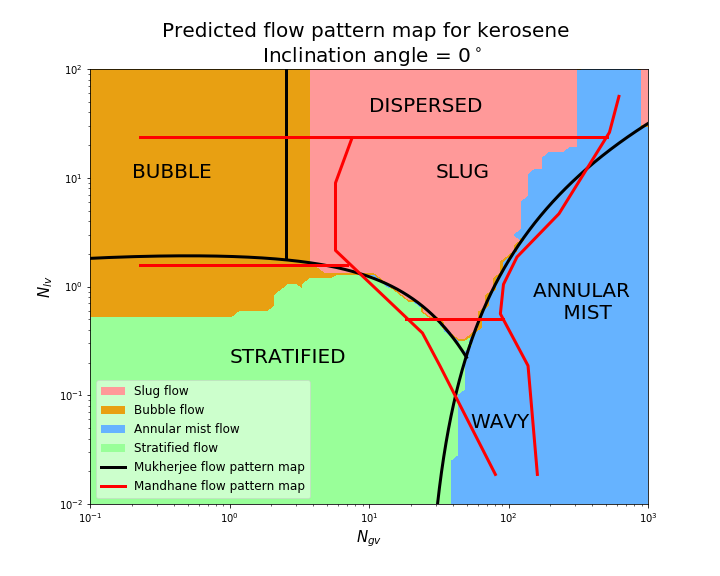}
	\caption{Flow pattern map for horizontal flow. Flow pattern obtained using ML algorithms are painted in different colours. In turn, boundaries from Mukherjee and Mandhane flow maps are displayed by lines}\label{fig17}
\end{figure}

After creation of the model for multiphase flow regime prediction it is possible to draw flow pattern maps. For example, in Fig.~\ref{fig17} the predicted map for horizontal flow of kerosene is represented. This map is generated by the Gradient Boosting algorithm. In this picture beside calculated zones of flow regimes, boundaries between classes obtained by Mukherjee \cite{mukherjee1979phd} and Mandhane \cite{mandhane1974flow} are drawn.
Using this figure one could identify that boundaries between flow patterns from Mukherjee and Mandhane maps are quite similar except for transition boundary between slug and bubble regimes. Boundaries predicted by the surrogate model almost restore Mukherjee map because the information about flow pattern in laboratory data is taken from Mukherjee experiments only. The greatest differences between the predicted map and Mukherjee (or Mandhane) map occur in ranges where data point are absent.

We will now consider the results of the third model which estimates pressure gradient in the segment. In this model, the pressure gradient is calculated using the equation for conservation of linear momentum (\ref{eq13}), as mentioned earlier. In this equation the friction factor is unknown. For estimation of this parameter, another one regression model is created. In this surrogate model division on horizontal, uphill and downhill flows does not apply because of small data points that belong to each group. The only one model for friction factor prediction is built that uses all data set in training, validation and testing stages. In this model, flow pattern feature is used among input parameters. The flow regime is encoded by creating several columns (each corresponding to a specific flow regime) and populating the data in the form 0 and 1. The method is referred to as one hot encoding. When a sample belongs to slug flow pattern, it will have 1 value in the column that is responsible for indication of an appurtenance to the slug flow class. The model for friction factor prediction is embedded in the calculation of the pressure gradient according to equation (\ref{eq13}), and, namely, the results of pressure gradient estimation will be discussed further. 

\begin{figure}[h!]
	\centering
	\includegraphics[width=0.5 \textwidth]{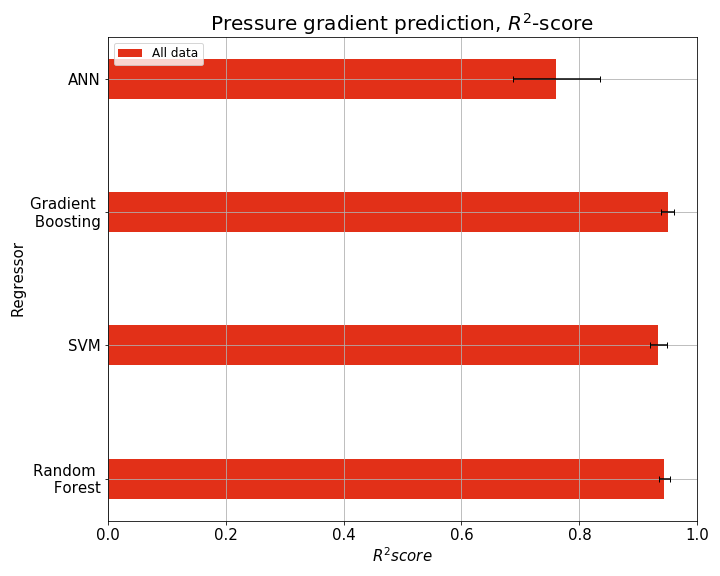}
	\caption{Histogram with $R^2$-scores in the model for pressure gradient prediction}\label{fig10}
\end{figure}

In Fig.~\ref{fig10} the comparative bar chart with different ML algorithms scores is plotted.

In this case, the Gradient Boosting algorithm provides the best score: $R^2 = 0.95 \pm 0.009$. From Fig.~\ref{fig10} it is possible to notice that three ML algorithms (Gradient Boosting, SVM and Random Forest) have approximately similar scores. For additional comparison of them, we represent computational time which each of these algorithms spends for execution the $M \times N$ ($M = 5, N = 20$) cross-validation procedure on the data set which consists of about 1300 data points. Gradient Boosting algorithm spends 15.41 seconds, Random Forest executes this operation within 24.78 seconds, and SVM algorithm performs this procedure during 47.06 seconds. The execution of cross-validation is carried out using notebook CPU Intel Core i7-8850H. From these values of time lengths, one could notice that the Gradient Boosting algorithm performs this procedure faster than SVM and Random Forest.

\begin{figure}[h!]
	\centering
	\includegraphics[width=0.5 \textwidth]{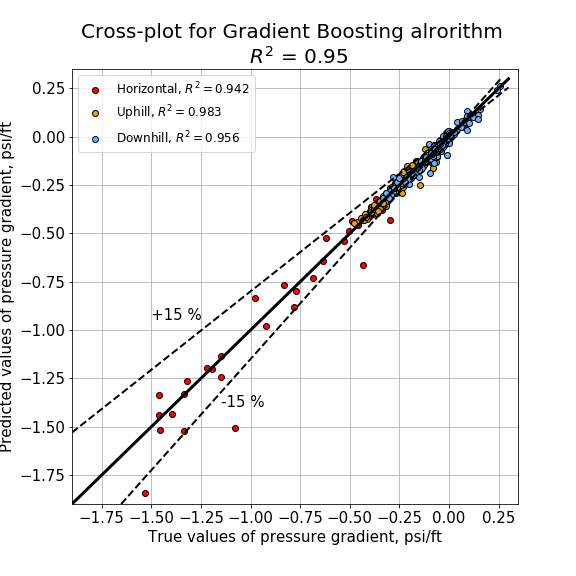}
	\caption{Cross-plot for Gradient Boosting algorithm in the pressure gradient model. Different colours are responsible for inclination angles}\label{fig11_1}
\end{figure}

\begin{figure}[h!]
	\centering
	\includegraphics[width=0.5 \textwidth]{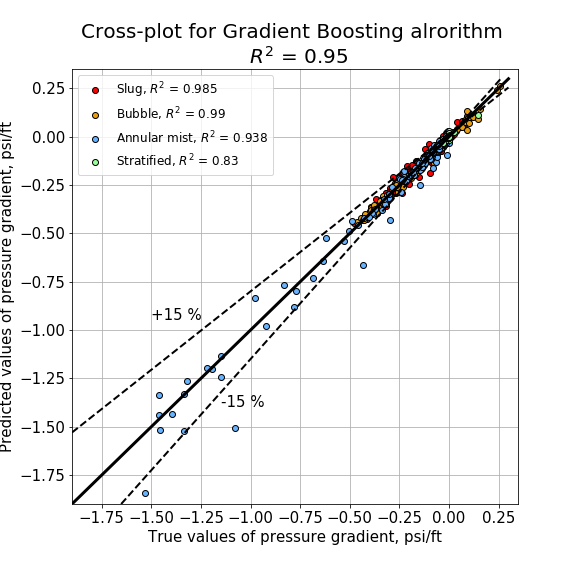}
	\caption{Cross-plot for Gradient Boosting algorithm in the pressure gradient model. Different colours are responsible for different flow regimes (observed in the experiment)}\label{fig11_2}
\end{figure}

\begin{figure}[h!]
	\centering
	\includegraphics[width=0.5 \textwidth]{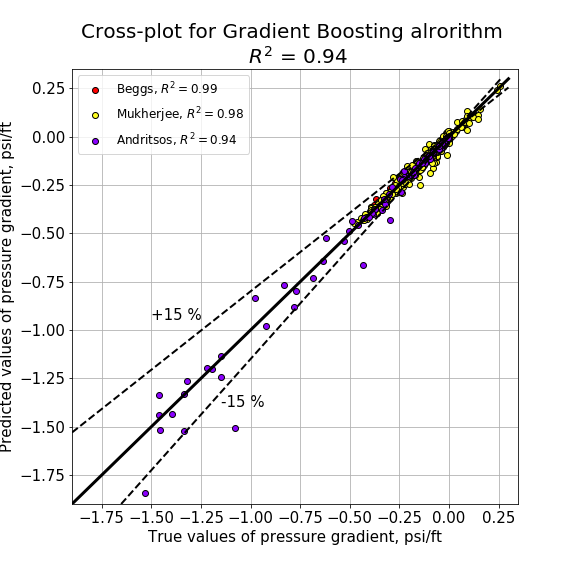}
	\caption{Cross-plot for Gradient Boosting algorithm in the pressure gradient model. Different colours are responsible for different experimentators}\label{fig11_3}
\end{figure}

Similarly to the liquid holdup model in this regression problem, it is also possible to analyze cross-plot. Different cross-plots with results of the best ML algorithm in the model - Gradient Boosting - are plotted in Fig.~\ref{fig11_1} - \ref{fig11_3}.

From Fig.~\ref{fig11_1} - \ref{fig11_3} one could observe that the number of outliers in this model is much less compared with the liquid holdup model. On the cross-plot Fig.~\ref{fig11_1} we marked points with different colours according to horizontal, upward and downward flows. From this plot, one could find out that beyond the 15\% error region data point belonging to horizontal and downhill flows are situated. From Fig.~\ref{fig11_2} it is possible to identify that in these cases flow regimes are annular mist and bubble. Finally, using Fig.~\ref{fig11_3} where we colour points according to data source, one could notice that the largest errors are obtained for measurements from Mukherjee and Andritsos datasets.

Having constructed the model for friction factor prediction using the Gradient Boosting algorithm, it is possible to observe what features are the most influential on this parameter. So, the list of features importance is the following:

\begin{enumerate}
\item Reynolds number
\item Froude number
\item Feature that indicates that sample belongs to the bubble flow pattern class
\item Viscosity number
\item Liquid velocity number
\item No-slip liquid holdup
\item Feature that indicates that sample belongs to the stratified flow pattern class
\item Liquid holdup
\item Relative roughness
\item Feature that indicates that sample belongs to the annular mist flow pattern class
\item Feature that indicates that sample belongs to the slug flow pattern class
\end{enumerate}

From this list, one could identify that flow pattern features and liquid holdup parameter have an impact on the prediction of friction factor and, as a result, on pressure gradient estimation.

Further, it is necessary to compare the predictive capability of the created model with empirical
correlations and mechanistic models. In order to perform this comparison, we use the same set of models as in the case of verification of liquid holdup model's results. The following coefficients of determination are obtained: Beggs \& Brill (0.863), Mukherjee \& Brill (0.921), TUFFP Unified (0.489), combination of Ansari and Xiao (0.912), Leda (0.9), OLGAS (0.87). As a result, Mukherjee \& Brill correlation and combination of Ansari and Xiao models give the best results. Further, we consider cross-plots for these two cases.

\begin{figure}[h!]
	\centering
	\includegraphics[width=0.5 \textwidth]{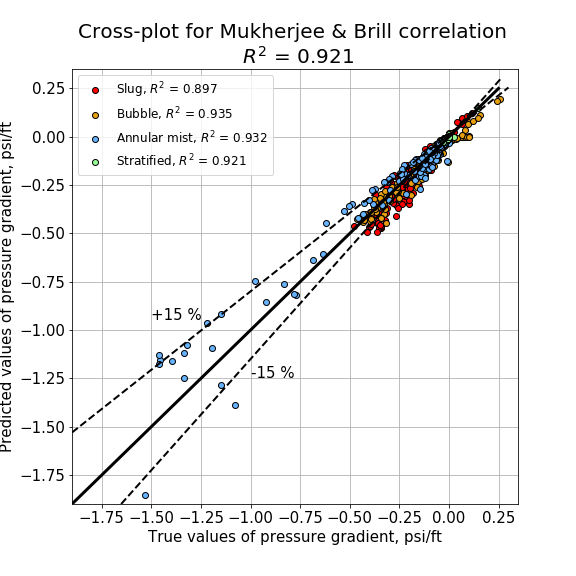}
	\caption{Cross-plot for result of Beggs \& Brill correlation}\label{fig12}
\end{figure}

In Fig.~\ref{fig12} pressure gradients obtained by Mukherjee \& Brill correlation are represented. Using this plot, it is possible to notice that this correlation predicts pressure gradients greater than $-0.5\frac{psi}{ft}$ a little bit worse in comparison with the proposed method. The certain number of these points lie outside region of 15 \% error, and the majority of them belongs to slug and annular mist regimes. The $R^2$-score is equal to 0.921 that is less than the Gradient Boosting algorithm provides. 

\begin{figure}[h!]
	\centering
	\includegraphics[width=0.5 \textwidth]{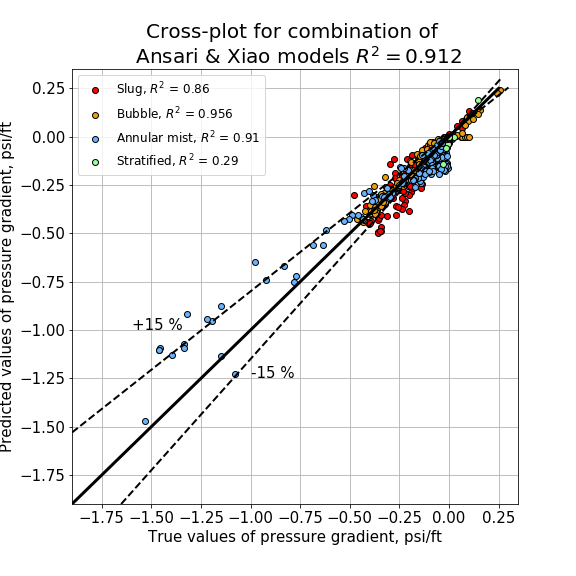}
	\caption{Cross-plot for result of Ansari \& Xiao combined model.}\label{fig12_1}
\end{figure}

In Fig.~\ref{fig12_1} we represent results of pressure gradients calculation using a combination of Ansari and Xiao mechanistic models. Using this chart, it is possible to see pressure gradients greater than $-0.5 \frac{psi}{ft}$ have wider spread around line ideal prediction. The majority of data points in this region that have error more than 15 \% belong to slug and annular most regimes.

So, the ML model created by using Gradient Boosting algorithm performs better than considered empirical and mechanistic models applied to lab database.  

In order to summarize results of created models, the Table \ref{tab2} with different scores (in the regression models: MSE, MAE and $R^2$; in classification: Accuracy and F1-macro scores) is written. In this table results of all applied machine learning algorithms in the case when all data set is used are represented.

\begin{table*}[h]
\begin{center}
\begin{tabular}{ | c | c | c | c |}
\hline
\multicolumn{4}{|c|}{Model for liquid holdup prediction}\\
\hline
Algorithm & $R^2$ & MSE & MAE \\ 
\hline
Random Forest & $0.924\pm 0.003$ & $0.005 \pm 0.0002$	& $0.038 \pm 0.0005$ \\ 
\hline
Gradient Boosting & $0.931 \pm 0.003$ &	$0.004 \pm	0.0002$ & $0.036 \pm	0.0004$ \\ 
\hline
SVM & $0.9 \pm 0.003$	& $0.006\pm 0.0002$	& $0.047 \pm	 0.0005$ \\ \hline
ANN & $0.897	\pm  0.004$	& $0.006	\pm 0.0002$	& $0.048\pm 0.0006$ \\ 
\hline
\end{tabular}
\begin{tabular}{ | c | c | c |}
\hline
\multicolumn{3}{|c|}{Model for flow pattern identification}\\
\hline
Algorithm & Accuracy & F1 - macro \\ 
\hline
Random Forest & $0.869\pm 0.003$ & $0.849 \pm 0.004$ \\ 
\hline
Gradient Boosting & $0.88 \pm 0.003$ &	$0.865 \pm	0.003$ \\ 
\hline
SVM & $0.864 \pm 0.003$	& $0.841\pm 0.003$\\ 
\hline
ANN & $0.85	\pm  0.003$	& $0.822 \pm 0.004$	\\ 
\hline
\end{tabular}
\begin{tabular}{ | c | c | c | c |}
\hline
\multicolumn{4}{|c|}{Model for pressure gradient prediction}\\
\hline
Algorithm & $R^2$ & MSE & MAE \\ 
\hline
Random Forest & $0.944\pm 0.009$ & $0.002 \pm 0.0004$	& $0.01 \pm 0.0004$ \\ 
\hline
Gradient Boosting & $0.95 \pm 0.009$ &	$0.002 \pm	0.0004$ & $0.01 \pm	0.0004$ \\ 
\hline
SVM & $0.934 \pm 0.014$	& $0.002\pm 0.0004$	& $0.012 \pm	 0.0005$ \\ \hline
ANN & $0.761	\pm  0.07$	& $0.008 \pm 0.0019$	& $0.021\pm 0.001$ \\ 
\hline
\end{tabular}
\caption{ML models results of all applied machine learning algorithms in the case of usage of all dataset.}
\label{tab2}
\end{center}
\end{table*}

\section{Sensitivity analysis}

This section is devoted to the elements of sensitivity analysis. Having constructed this set of surrogate models above, it is necessary to evaluate their behaviour when input parameters are changed. To perform this evaluation, we select the base set of values of input parameters such as superficial gas and liquid velocities, gas and liquid densities, viscosities, gas-liquid surface tension, diameter of the pipe segment and its inclination angle. Each of these parameters is varied within its range of variation, while other input parameters are fixed equal to their base values. 

This procedure is carried out for the liquid holdup prediction and estimation of the pressure gradient. Using the values of these output parameters, we construct tornado diagrams. Several plots with dependences of pressure gradient on crucial features such as superficial velocities, diameter and inclination angle are constructed. On these plots, we compare results of proposed surrogate models with Beggs \& Brill correlation.  

We consider three different cases of a base set of input parameters. In the first set inclination angle is zero, in the second one it is greater than zero, and in the last one is less than zero. This choice is made because we construct different surrogate models for horizontal, uphill and downhill flows for liquid holdup value prediction and flow pattern identification. Values of other input parameters in base sets are taken equal to average values of these parameters from the lab data set (in the first case we consider all data, in the second one data points related to the uphill flow and in the third one data responsible for the downhill flow). Each of input parameters is varied in value range that is contained in the corresponding range presented in the lab database besides diameter which variation range is chosen more extensive than in the lab database.

Let us begin with the first set of base values of input parameters: $\theta = 0^{\circ}, v_{sg} = 44.075 \frac{ft}{sec}, v_{sl} = 2.13 \frac{ft}{sec}, \rho_l = 58.03 \frac{lb}{ft^3}, \rho_g = 0.2048 \frac{lb}{ft^3}, \mu_l = 6.197 cP, \mu_g = 0.0211 cP, \sigma = 49.86 \frac{Dynes}{cm}, d = 0.125 ft$. In this case liquid holdup value - $\alpha_l = 0.153$ and pressure gradient  - $\frac{dp}{dl} = -0.158 \frac{psi}{ft}$.

\begin{figure}[h!]
	\centering
	\includegraphics[width=0.5 \textwidth]{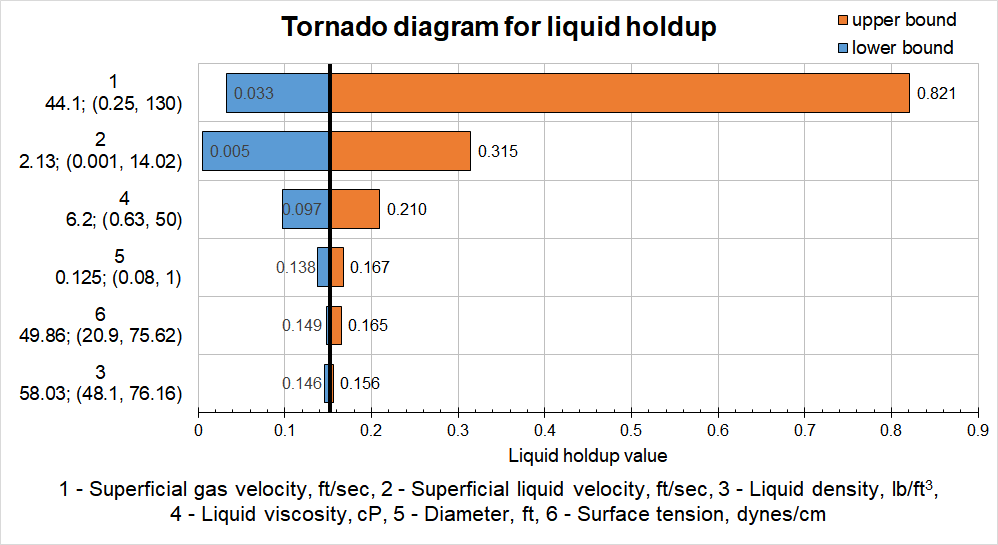}
	\caption{Tornado diagram with variations of liquid holdup value when base case belongs to horizontal flow. On the left of the chart, under the parameter's sequence number the base value and its variation range are written}\label{fig28_1}
\end{figure}

\begin{figure}[h!]
	\centering
	\includegraphics[width=0.5 \textwidth]{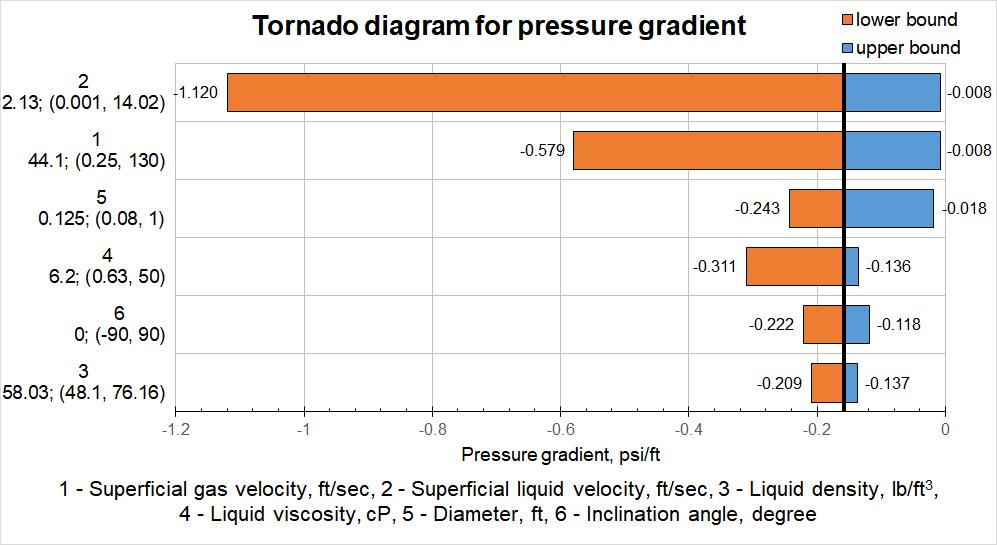}
	\caption{Tornado diagram with variations of pressure gradient value when base case belongs to horizontal flow. On the left of the chart, under the parameter's sequence number the base value and its variation range are written}\label{fig28_2}
\end{figure}

In Fig.~\ref{fig28_1} and \ref{fig28_2} the tornado diagrams for liquid holdup and pressure gradient variations are represented. We represent the six most influential parameters on the output value. Variation ranges of input parameters are also written in these plots. From these diagrams, one could find out that superficial velocities, liquid viscosity and pipe diameter have the greatest influence not only on liquid holdup but also on pressure gradient.  

Further, we move on the second set of base values. In this set, the inclination angle is greater than zero and equals to $50^{\circ}$: $\theta = 50^{\circ}, v_{sg} = 23.83 \frac{ft}{sec}, v_{sl} = 3.81 \frac{ft}{sec}, \rho_l = 52.52 \frac{lb}{ft^3}, \rho_g = 0.278 \frac{lb}{ft^3}, \mu_l = 3.4 cP, \mu_g = 0.0211 cP, \sigma = 34 \frac{Dynes}{cm}, d = 0.12 ft$. Values of liquid holdup and pressure gradient are the following: $\alpha_l = 0.275, \frac{dp}{dl} = -0.217 \frac{psi}{ft}$.

\begin{figure}[h!]
	\centering
	\includegraphics[width=0.5 \textwidth]{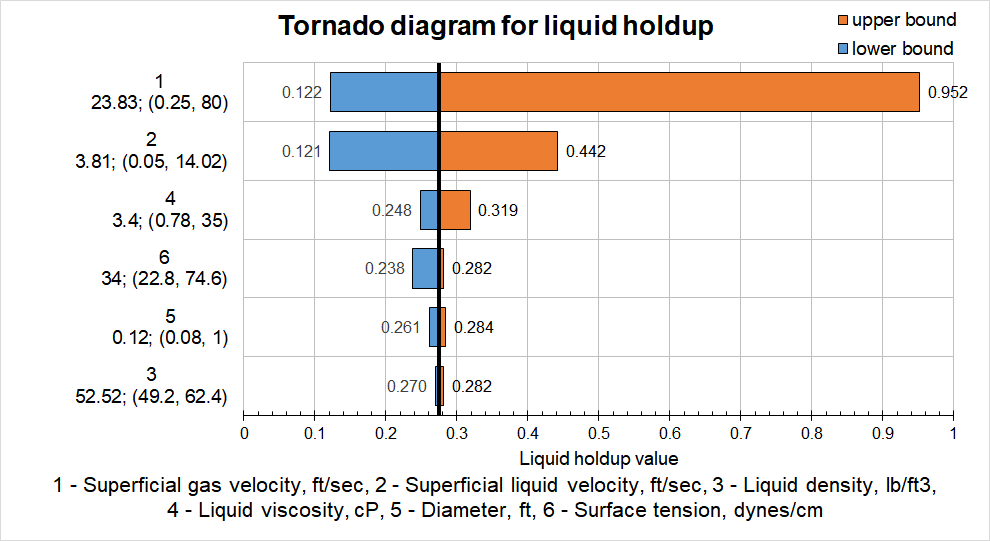}
	\caption{Tornado diagram with variations of liquid holdup value when base case belongs to uphill flow. On the left of the chart, under the parameter's sequence number the base value and its variation range are written}\label{fig28_3}
\end{figure}

\begin{figure}[h!]
	\centering
	\includegraphics[width=0.5 \textwidth]{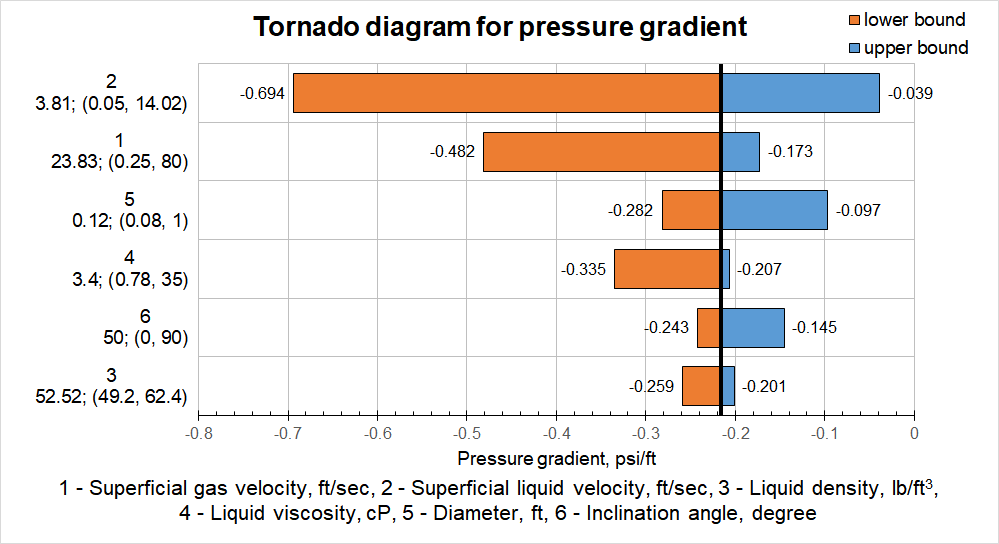}
	\caption{Tornado diagram with variations of pressure gradient value when base case belongs to uphill flow. On the left of the chart, under the parameter's sequence number the base value and its variation range are written}\label{fig28_4}
\end{figure}

In this case, the tornado charts for the liquid holdup and the pressure gradient are also constructed, and they are represented in Fig.~\ref{fig28_3} and \ref{fig28_4}. From these figures it is noticeable that the most important features for liquid holdup prediction are superficial velocities and liquid viscosity; for pressure gradient prediction these features are superficial velocities, tube diameter and liquid viscosity. 

Now we turn to the final set of base variables with inclination angle less than zero and equals to $-50^{\circ}$: $\theta = -50^{\circ}, v_{sg} = 29.31 \frac{ft}{sec}, v_{sl} = 3.1 \frac{ft}{sec}, \rho_l = 52.57 \frac{lb}{ft^3}, \rho_g = 0.26 \frac{lb}{ft^3}, \mu_l = 1 cP, \mu_g = 0.024 cP, \sigma = 34 \frac{Dynes}{cm}, d = 0.12 ft$. Liquid holdup value for this set is equal to $\alpha_l = 0.225$ and pressure gradient is the following: $\frac{dp}{dl} = -0.066 \frac{psi}{ft}$.

\begin{figure}[h!]
	\centering
	\includegraphics[width=0.5 \textwidth]{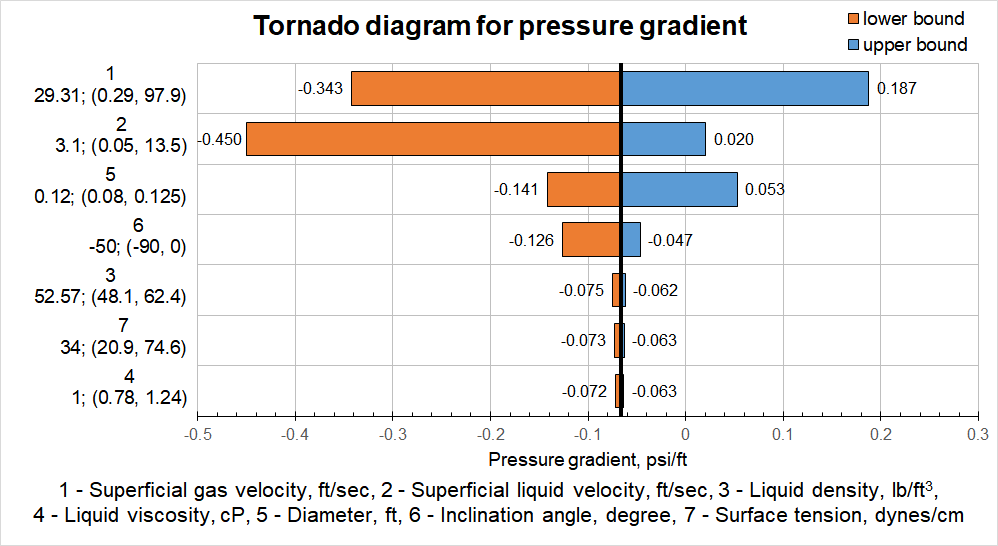}
	\caption{Tornado diagram with variations of liquid holdup value when base case belongs to downhill flow. On the left of the chart, under the parameter's sequence number the base value and its variation range are written}\label{fig28_5}
\end{figure}

\begin{figure}[h!]
	\centering
	\includegraphics[width=0.5 \textwidth]{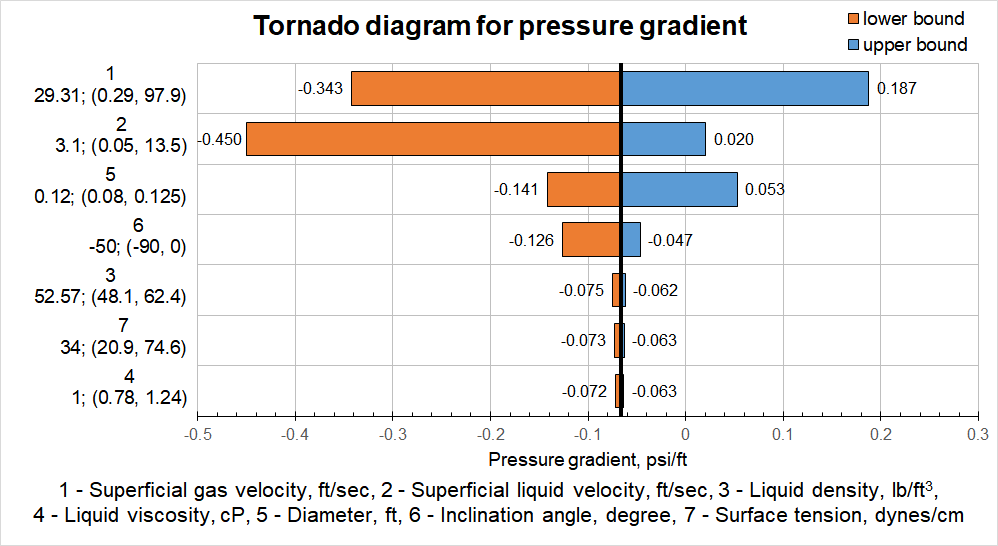}
	\caption{Tornado diagram with variations of pressure gradient value when base case belongs to downhill flow. On the left of the chart, under the parameter's sequence number the base value and its variation range are written}\label{fig28_6}
\end{figure}

In Fig.~\ref{fig28_5} and \ref{fig28_6} tornado diagrams with the seven most substantial features are represented. Using this set of base values, we obtain that both liquid holdup and pressure gradient values significantly depends on superficial velocities, diameter of the tube segment and its inclination angle.

\begin{figure}[h!]
	\centering
	\includegraphics[width=0.4 \textwidth]{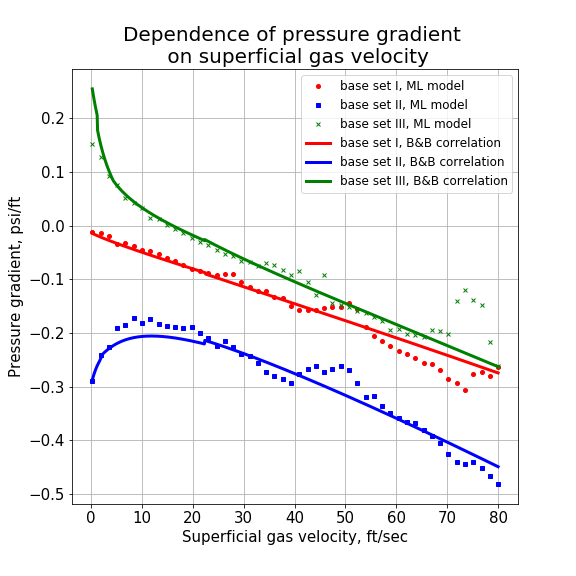}
	\caption{Dependence of pressure gradient on superficial gas velocity for all considered base sets of input parameters obtained by using propose method and Beggs \& Brill correlation}\label{fig29_1}
\end{figure}

\begin{figure}[h!]
	\centering
	\includegraphics[width=0.4 \textwidth]{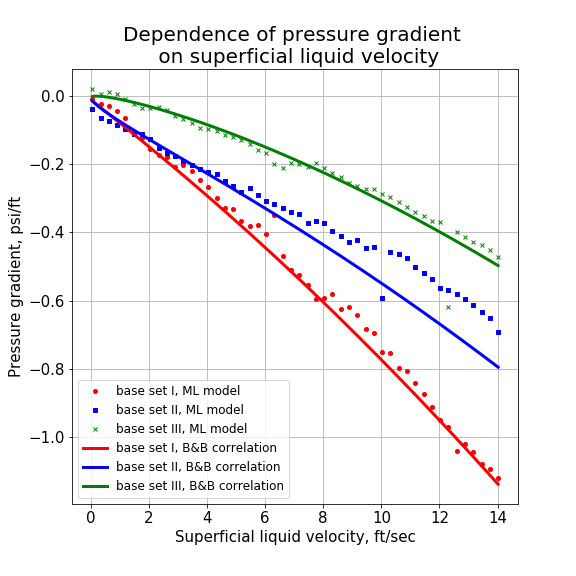}
	\caption{Dependence of pressure gradient on superficial liquid velocity for all considered base sets of input parameters obtained by using propose method and Beggs \& Brill correlation}\label{fig29_2}
\end{figure}

\begin{figure}[h!]
	\centering
	\includegraphics[width=0.4 \textwidth]{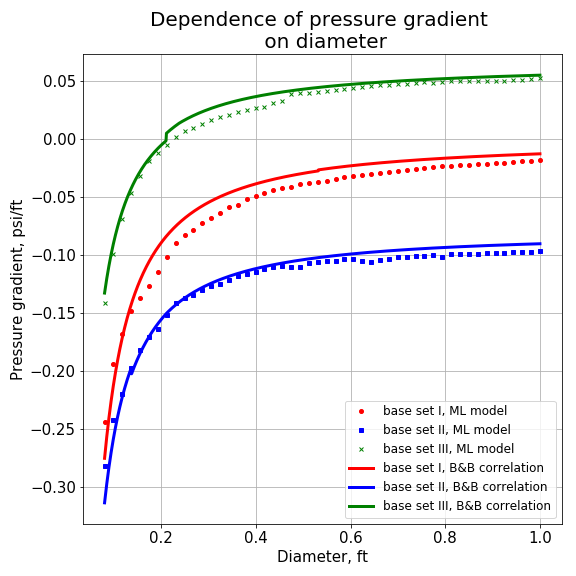}
	\caption{Dependence of pressure gradient on pipe diameter for all considered base sets of input parameters obtained by using propose method and Beggs \& Brill correlation}\label{fig29_3}
\end{figure}

\begin{figure}[h!]
	\centering
	\includegraphics[width=0.4 \textwidth]{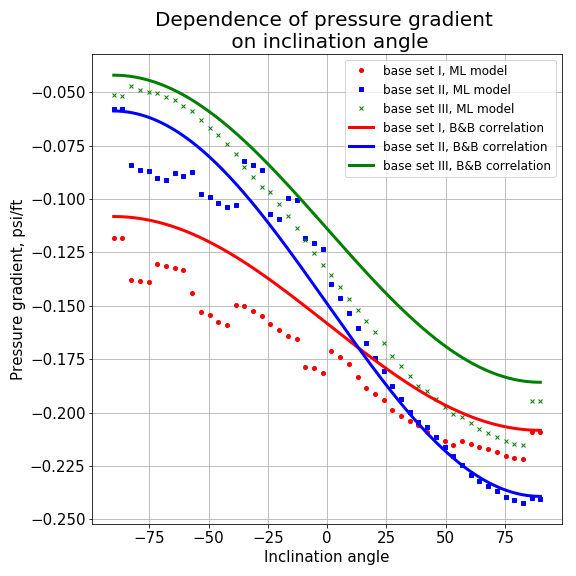}
	\caption{Dependence of pressure gradient on pipe inclination angle for all considered base sets of input parameters obtained by using propose method and Beggs \& Brill correlation}\label{fig29_4}
\end{figure}

At the end of this section we represent graphs with dependencies of pressure gradient on superficial velocities, pipe diameter and inclination angle obtained with the help of surrogate models and Beggs \& Brill correlation (Fig.~\ref{fig29_1} - \ref{fig29_4}). Using these plots one could find out that the proposed method gives profiles that are quite similar to dependences obtained by using considered correlation. The main differences could be observed on the chart with the dependence of pressure gradient on inclination angle (Fig.~\ref{fig29_4}).

\section{Case study}
\label{Case_study_section}
Here we will represent the validation results of the constructed method for pressure drop calculation on field data. Three field data sets with measurements on production wells and pipelines are taken from \cite{asheim1986mona}. These data sets include flow rates, bottomhole and wellhead pressures and temperatures, flow length, densities of oil and gas at standard conditions, average inclination angles of the pipes.

\begin{figure}[h!]
	\centering
	\includegraphics[width=0.45 \textwidth]{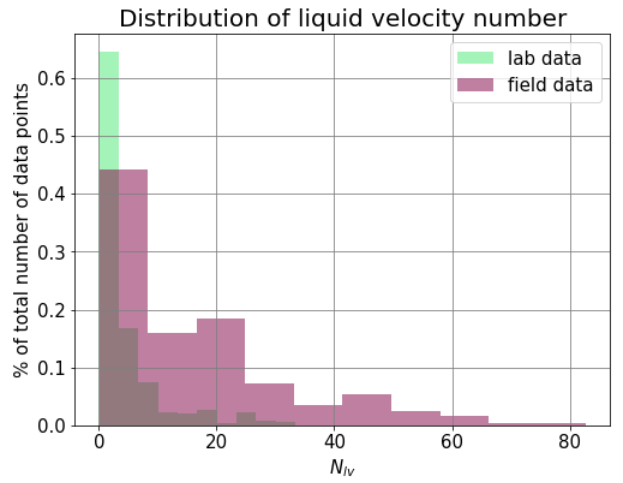}
	\caption{Bar chart with comparison of ranges of liquid velocity number in laboratory and field data}
	\label{fig21}
\end{figure}

\begin{figure}[h!]
	\centering
	\includegraphics[width=0.45 \textwidth]{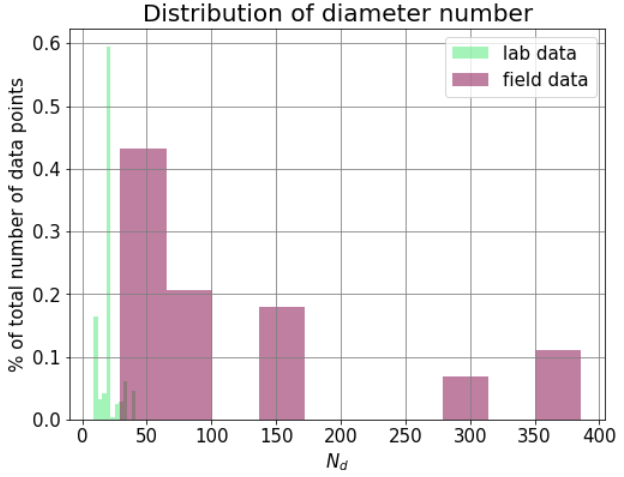}
	\caption{Bar chart with comparison of ranges of diameter number in laboratory and field data}
	\label{fig22}
\end{figure}

\begin{figure}[h!]
	\centering
	\includegraphics[width=0.45 \textwidth]{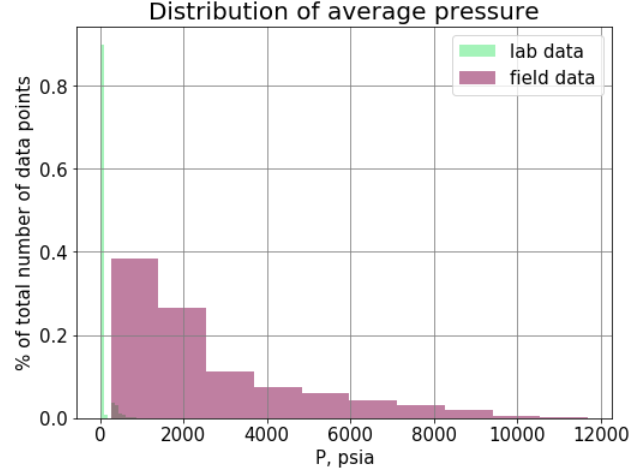}
	\caption{Bar chart with comparison of ranges of average pressure in laboratory and field data}
	\label{fig23}
\end{figure}

\begin{figure}[h!]
	\centering
	\includegraphics[width=0.45 \textwidth]{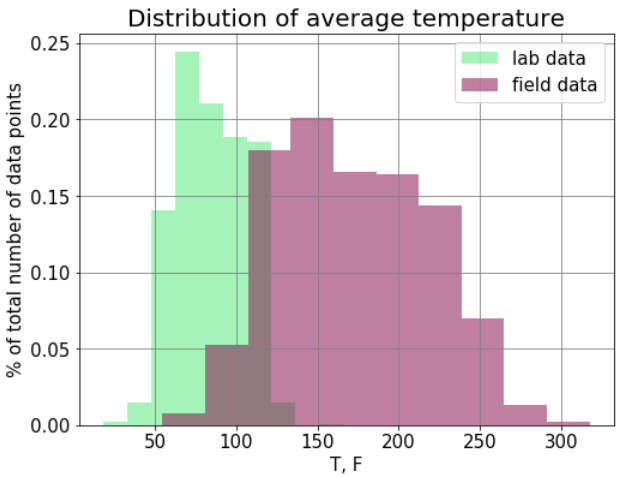}
	\caption{Bar chart with comparison of ranges of average temperature in laboratory and field data}
	\label{fig24}
\end{figure}

\begin{figure}[h!]
	\centering
	\includegraphics[width=0.45 \textwidth]{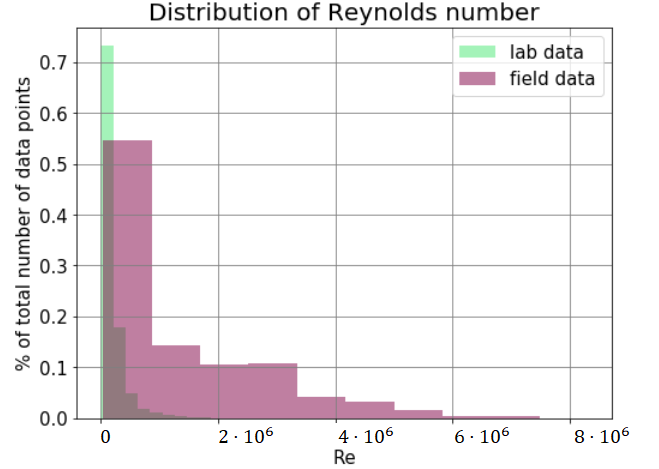}
	\caption{Bar chart with comparison of ranges of Reynolds number in laboratory and field data}
	\label{fig25}
\end{figure}

First of all, it is necessary to compare data sets composed of laboratory experiments and field measurements. In order to calculate from field data the same parameters of multiphase flow in wells and pipelines as in the laboratory database, namely, velocities, densities and viscosities of gas and liquid phases, the method from the article \cite{li2014combined} is applied. The pipe is divided into segments, and multiphase flow correlations are applied (in the present study, Beggs \& Brill and Mukherjee \& Brill \cite{mukherjee1979phd} correlations are chosen). In other words, the field measurements are available only at the inlet (outlet) of the tube, while correlation allows one to populate these data along the pipe. We apply both correlations for each of the field data samples and choose this one that yields a closer value of bottomhole pressure to the measured value. Synthetic data (the result of computation) is assumed acceptable if the difference between calculated bottomhole pressure and the real value within 5 \% accuracy. Then, synthetic data is populated in the segments along the well (pipeline). Among all parameters that could be utilized as input data in surrogate models (inclination angle, gas and liquid velocity numbers, viscosity number, diameter number, no-slip liquid holdup, Reynolds number, Froude number, average pressure and temperature) the following ones have different value ranges: liquid velocity number, diameter number, Reynolds number. In Fig.~\ref{fig21} - \ref{fig25} the bar charts with comparison of this parameters are depicted.

From Fig.~\ref{fig22}, \ref{fig23} one could reveal that ranges of diameter number and average pressure in the case of field and lab data almost do not overlap. In this regard, during the construction of ML models, these parameters should have been excluded because machine learning algorithms could not be applied outside the training ranges. Ranges of average temperature parameter (Fig.~\ref{fig24}) in the cases of lab and field data sets also slightly overlap; that is why this feature is not included in training parameters. However, the temperature has already been utilized during the calculation of density, viscosity and surface tension. So, it is not necessary to take into account it once again. Since the ranges of liquid velocity number and Reynolds number in the case of lab data and field data only partially overlap, there might be potential errors during application of constructed model on the field data.

Let us consider each data set in order. The first data set consists of production tests on inclined wells from The Forties field (The United Kingdom). The characteristic feature of the produced fluid in this field is relatively low gas content. Computations show that the flow is a single phase along the large part of the flow length on each well. At the same time, in the two-phase region slug and bubble flow regimes are encountered.

\begin{figure}[h!]
	\centering
	\includegraphics[width=0.5 \textwidth]{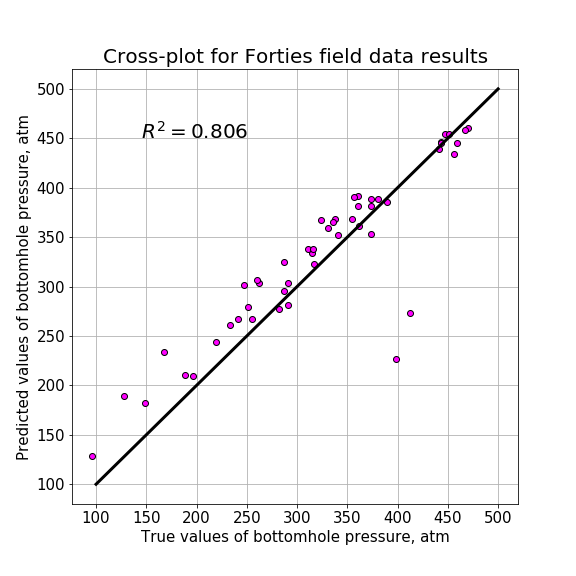}
	\caption{Cross-plot with result of bottom hole pressure calculation on Forties production wells}
	\label{fig18}
\end{figure}

In Fig.~\ref{fig18} the cross-plot with obtained results of bottom hole pressure calculation is depicted.

The proposed method yields coefficient of determination equal to $R^2 = 0.806$ on this data set. This result is a little bit worse compared to the application of Beggs \& Brill correlation ($R^2 = 0.82$). However, the created model performs a little bit better on this data set than the following set of correlations and mechanistic models: Mukherjee \& Brill correlation ($R^2 = 0.79$), Ansari \& Xiao combined model ($R^2 = 0.66$), TUFFP Unified ($R^2 = 0.795$), Leda Flow point model ($R^2 = 0.793$).  

Let us move on to the results of the application of the constructed model concerning the second data set. It consists of production tests on inclined wells from Ekofisk area (Norway). These wells produce light oil and have a smaller diameter of tubing than ones from Forties field.

\begin{figure}[h!]
	\centering
	\includegraphics[width=0.45 \textwidth]{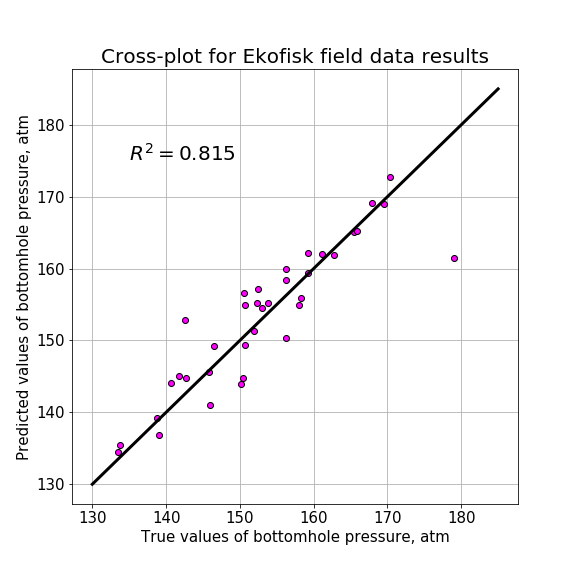}
	\caption{Cross-plot with result of bottom hole pressure calculation on Ekofisk production wells}
	\label{fig19}
\end{figure}

In Fig.~\ref{fig19} one could observe the results of application the constructed method for this data set. The model provides $R^2 = 0.815$ that is better than models with Beggs \& Brill correlation ($R^2 = 0.65$), Ansari \& Xiao combined model ($R^2 = 0.78$) and TUFFP Unified ($R^2 = 0.73$). However, the score obtained by using surrogate models is slightly worse than Mukherjee \& Brill ($R^2 = 0.823$) and Leda Flow point model ($R^2 = 0.821$) allow obtaining.

The final data set is composed of test data for large diameter flowlines at Prudhoe Bay field (USA). These flow lines are nearly horizontal with a small inclination angle.

\begin{figure}[h!]
	\centering
	\includegraphics[width=0.45 \textwidth]{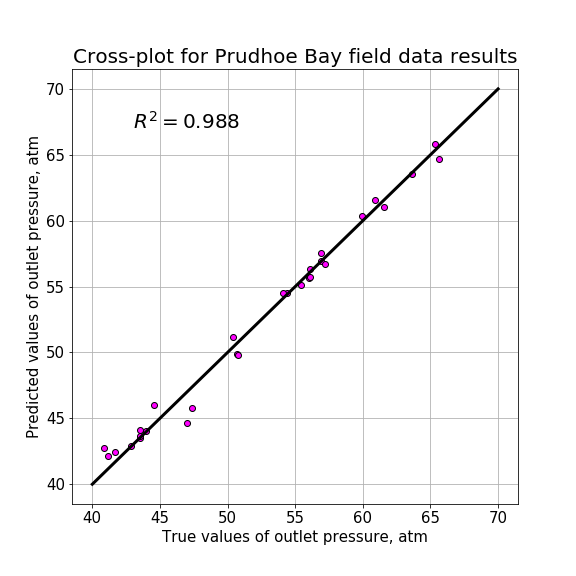}
	\caption{Cross-plot with result of bottom hole pressure calculation on Prudhoe Bay flowlines}
	\label{fig20}
\end{figure}

The cross-plot with calculated outlet pressures on the flowlines is represented in Fig.~\ref{fig20}. The obtained coefficient of determination on this data set is relatively high ($R^2 = 0.99$). In turn, Beggs \& Brill correlation gives $R^2 = 0.98$ that is also very high result but a little worse. In contrast, the other considered methods provide even lower scores: Mukherjee \& Brill correlation ($R^2 = 0.94$), Ansari \& Xiao combined model ($R^2 = 0.78$), TUFFP Unified ($R^2 = 0.47$), Leda Flow point model ($R^2 = 0.79$).

\section{Conclusions}

In this paper, a new methodology for pressure gradient calculation on a pipe segment is proposed. The method consists of three surrogate models nested in each other: the first model predicts the value of liquid holdup. The best $R^2$-score (0.93) for this model was achieved by using the Gradient Boosting algorithm. The second model is focused on the flow pattern determination using the output result of the first model. The best accuracy score (88.1 \%) by this model is provided by the Gradient Boosting algorithm. The third model predicts pressure gradient based on liquid holdup and flow regime determined in the first two steps of the workflow. In the case of the third model, the Gradient Boosting method performed the best ($R^2 = 0.95$). Further, all of these surrogate models are applied in the iterative algorithm for calculating the outlet (inlet) pressure in the pipe based on the inlet (outlet) pressure and other parameters. The distinguishing feature of the proposed method is its wider applicability range in terms of the input parameters, which is a result of the application of the ML algorithms, their training on a larger lab database, and a complex approach for pressure gradient calculation. The combined method proposed in this study provides results with comparable or higher accuracy for liquid holdup and pressure gradient predictions on the considered lab data set as compared with several multiphase flow correlations and mechanistic models. For example, OLGAS model yields a coefficient of determination $R^2 = 0.892$ for liquid holdup calculation. In case of pressure gradient calculations, Mukherjee \& Brill correlation provides a coefficient of determination, which equals to 0.92. In addition, the proposed method was tested on three field data sets. 

It is important to note that we carry out calculations for three field datasets using the ML algorithms trained on the lab data without any retraining. This validation shows that utilizing the constructed method for calculating the pressure distribution along the pipe it is possible to obtain the following coefficients of determination: $R^2 = 0.806, 0.815, 0.99$. In contrast, by applying correlations and mechanistic models to these datasets, it is possible to draw the following best scores: $R^2 = 0.82$ (by using Beggs and Brill correlation), $R^2 = 0.823$ (Mukherjee and Brill correlation) and $R^2 = 0.98$ (Beggs and Brill correlation). As a result, we could obtain pressure values (e.g., bottom hole pressure of the well) using ML model with a comparable or even higher accuracy as the considered multiphase flow correlation and mechanistic models provide. Thus, ML algorithms can be scaled from the lab to the field conditions.

For future studies, we see several ways to improve the predictive capability of the proposed model. First, the extension of the lab data set can help expand the applicability margins. It could be either experimental or synthetic data calculated using steady-state multiphase flow simulators such as PipeSim, OLGAS, and Leda Flow point model. Second, an accurate description of the temperature effects should be introduced. In the present paper, the temperature is linearly interpolated between the boundaries. However, it is better to calculate the change of entropy and heat transfer along the pipe for more precise calculation of temperature profile. 

The workflow proposed so far covers the class of steady-state multiphase flows in vertical and inclined wells and pipelines. This model can be used as an engineering tool for calculating the bottomhole pressure based on surface pressure measurements at the well head, for example during flowback of a multistage fractured well, when it is essential to control the drawdown to make sure it is not excessive, to avoid undesired geomechanics phenomena (proppant flowback, tensile rock failure, fracture pinch-out).

Another significant domain of applicability for such method will be to define the limits of applicability of this steady-state approach for dynamic conditions for fractured horizontal well production in order to avoid pressure fluctuation in front of the fracture. This should fix the safe operating envelope for well exploitation. Also, to qualitatively define such limits we can use non-parametric conformal measures to estimate predictive confidence regions \cite{VovkConformal2014,ConformalKRR2016}. 

For future work, one may also consider incorporating machine learning algorithms into a mechanistic model. For example, in a two-fluid stratified flow model a machine learning algorithm could be used to compute the interfacial friction factor, which is an important closure relation. The benefit is the number of input parameters should decrease as the physics is captured by the model formulation and not merely dimensionless quantities.


\section*{Acknowledgements}
{The authors are grateful to the management of LLC "Gazpromneft-STC" for supporting this initiative. The work received partial financial support from the Ministry of Education and Science of the Russian Federation (project №14.581.21.0027, unique identifier RFMEFI58117X0027).

Startup funds of Skolkovo Institute of Science and Technology are gratefully acknowledged by A.A. Osiptsov.}
\bibliographystyle{elsarticle-num}
\bibliography{References}

\begin{thebibliography}{10}
\expandafter\ifx\csname url\endcsname\relax
  \def\url#1{\texttt{#1}}\fi
\expandafter\ifx\csname urlprefix\endcsname\relax\def\urlprefix{URL }\fi
\expandafter\ifx\csname href\endcsname\relax
  \def\href#1#2{#2} \def\path#1{#1}\fi

\bibitem{ladva2000multiphase}
H.~Ladva, P.~Tardy, P.~Howard, V.~Dussan, et~al., Multiphase flow and drilling
  fluid filtrate effects on the onset of production, in: SPE International
  Symposium on Formation Damage Control, Society of Petroleum Engineers, 2000.

\bibitem{sun2018multiphase}
B.~Sun, Y.~Guo, W.~Sun, Y.~Gao, H.~Li, Z.~Wang, H.~Zhang, Multiphase flow
  behavior for acid-gas mixture and drilling fluid flow in vertical wellbore,
  Journal of Petroleum Science and Engineering 165 (2018) 388--396.

\bibitem{osiptsov2017review}
A.~A. Osiptsov, Fluid mechanics of hydraulic fracturing: a review, Journal of
  Petroleum Science and Engineering 156 (2017) 513--535.

\bibitem{bratland2010pipe}
O.~Bratland, Pipe flow 2: Multi-phase flow assurance, Ove Bratland.

\bibitem{osman2004artificial}
E.-S. Osman, Artificial neural network models for identifying flow regimes and
  predicting liquid holdup in horizontal multiphase flow, Old Production \&
  Facilities 19~(1) (2004) 33--40.

\bibitem{li2014multiphase}
X.~Li, J.~L. Miskimins, R.~P. Sutton, B.~T. Hoffman, et~al., Multiphase flow
  pattern recognition in horizontal and upward gas-liquid flow using support
  vector machine models, in: SPE Annual Technical Conference and Exhibition,
  Society of Petroleum Engineers, 2014.

\bibitem{popa2015fuzzynistic}
F.~Popa, S.~Dursun, B.~Houchens, et~al., Fuzzynistic models for multiphase flow
  pattern identification, in: SPE Annual Technical Conference and Exhibition,
  Society of Petroleum Engineers, 2015.

\bibitem{shippen2002neural}
M.~E. Shippen, S.~L. Scott, et~al., A neural network model for prediction of
  liquid holdup in two-phase horizontal flow, in: SPE Annual Technical
  Conference and Exhibition, Society of Petroleum Engineers, 2002.

\bibitem{aziz1972pressure}
K.~Aziz, G.~W. Govier, et~al., Pressure drop in wells producing oil and gas,
  Journal of Canadian Petroleum Technology 11~(03).

\bibitem{beggs1973study}
D.~H. Beggs, J.~P. Brill, et~al., A study of two-phase flow in inclined pipes,
  Journal of Petroleum technology 25~(05) (1973) 607--617.

\bibitem{mukherjee1983liquid}
H.~Mukherjee, J.~P. Brill, et~al., Liquid holdup correlations for inclined
  two-phase flow, Journal of Petroleum Technology 35~(05) (1983) 1--003.

\bibitem{hasan1986study}
A.~Hasan, C.~Kabir, et~al., A study of multiphase flow behavior in vertical oil
  wells: Part i-theoretical treatment, in: SPE California Regional Meeting,
  Society of Petroleum Engineers, 1986.

\bibitem{ansari1990comprehensive}
A.~Ansari, N.~Sylvester, O.~Shoham, J.~Brill, et~al., A comprehensive
  mechanistic model for upward two-phase flow in wellbores, in: SPE Annual
  Technical Conference and Exhibition, Society of Petroleum Engineers, 1990.

\bibitem{zhang2003unified_1}
H.-Q. Zhang, Q.~Wang, C.~Sarica, J.~P. Brill, A unified mechanistic model for
  slug liquid holdup and transition between slug and dispersed bubble flows,
  International journal of multiphase flow 29~(1) (2003) 97--107.

\bibitem{zhang2003unified_2}
H.-Q. Zhang, Q.~Wang, C.~Sarica, J.~P. Brill, Unified model for gas-liquid pipe
  flow via slug dynamics—part 1: model development, Journal of energy
  resources technology 125~(4) (2003) 266--273.

\bibitem{danielson2005leda}
T.~J. Danielson, K.~M. Bansal, R.~Hansen, E.~Leporcher, et~al., Leda: the next
  multiphase flow performance simulator, in: 12th International Conference on
  Multiphase Production Technology, BHR Group, 2005.

\bibitem{bendiksen1991dynamic}
K.~H. Bendiksen, D.~Maines, R.~Moe, S.~Nuland, et~al., The dynamic two-fluid
  model olga: Theory and application, SPE production engineering 6~(02) (1991)
  171--180.

\bibitem{osman2005artificial}
E.-S.~A. Osman, M.~A. Ayoub, M.~A. Aggour, et~al., An artificial neural network
  model for predicting bottomhole flowing pressure in vertical multiphase flow,
  in: SPE Middle East Oil and Gas Show and Conference, Society of Petroleum
  Engineers, 2005.

\bibitem{li2014combined}
X.~Li, J.~Miskimins, B.~T. Hoffman, et~al., A combined bottom-hole pressure
  calculation procedure using multiphase correlations and artificial neural
  network models, in: SPE Annual Technical Conference and Exhibition, Society
  of Petroleum Engineers, 2014.

\bibitem{kanin2018method}
E.~Kanin, A.~Vainshtein, A.~Osiptsov, E.~Burnaev, The method of calculation the
  pressure gradient in multiphase flow in the pipe segment based on the machine
  learning algorithms, in: IOP Conference Series: Earth and Environmental
  Science, Vol. 193, IOP Publishing, 2018, p. 012028.

\bibitem{GTApprox2016}
M.~Belyaev, E.~Burnaev, E.~Kapushev, M.~Panov, P.~Prikhodko, D.~Vetrov,
  D.~Yarotsky, Gtapprox: Surrogate modeling for industrial design, Advances in
  Engineering Software 102 (2016) 29 -- 39.
\newblock \href
  {http://dx.doi.org/https://doi.org/10.1016/j.advengsoft.2016.09.001}
  {\path{doi:https://doi.org/10.1016/j.advengsoft.2016.09.001}}.

\bibitem{brill1999multiphase}
J.~P. Brill, H.~K. Mukherjee, Multiphase flow in wells, Vol.~17, Society of
  Petroleum Engineers, 1999.

\bibitem{duns1963vertical}
H.~Duns~Jr, N.~Ros, et~al., Vertical flow of gas and liquid mixtures in wells,
  in: 6th World Petroleum Congress, World Petroleum Congress, 1963.

\bibitem{breiman2001random}
L.~Breiman, Random forests, Machine learning 45~(1) (2001) 5--32.

\bibitem{scikit-learn}
F.~Pedregosa, G.~Varoquaux, A.~Gramfort, V.~Michel, B.~Thirion, O.~Grisel,
  M.~Blondel, P.~Prettenhofer, R.~Weiss, V.~Dubourg, J.~Vanderplas, A.~Passos,
  D.~Cournapeau, M.~Brucher, M.~Perrot, E.~Duchesnay, Scikit-learn: Machine
  learning in {P}ython, Journal of Machine Learning Research 12 (2011)
  2825--2830.

\bibitem{friedman1999greedy}
J.~Friedman, Greedy function approximation: A gradient boosting machine
  http://www. salford-systems. com/doc, GreedyFuncApproxSS. pdf.

\bibitem{Chen:2016:XST:2939672.2939785}
T.~Chen, C.~Guestrin,
  \href{http://doi.acm.org/10.1145/2939672.2939785}{{XGBoost}: A scalable tree
  boosting system}, in: Proceedings of the 22nd ACM SIGKDD International
  Conference on Knowledge Discovery and Data Mining, KDD '16, ACM, New York,
  NY, USA, 2016, pp. 785--794.
\newblock \href {http://dx.doi.org/10.1145/2939672.2939785}
  {\path{doi:10.1145/2939672.2939785}}.
\newline\urlprefix\url{http://doi.acm.org/10.1145/2939672.2939785}

\bibitem{makhotin2019gradient}
I.~Makhotin, D.~Koroteev, E.~Burnaev, Gradient boosting to boost the efficiency
  of hydraulic fracturing, arXiv preprint arXiv:1902.02223.

\bibitem{ignatov2018tree}
D.~I. Ignatov, K.~Sinkov, P.~Spesivtsev, I.~Vrabie, V.~Zyuzin, Tree-based
  ensembles for predicting the bottomhole pressure of oil and gas well flows,
  in: International Conference on Analysis of Images, Social Networks and
  Texts, Springer, 2018, pp. 221--233.

\bibitem{cortes1995support}
C.~Cortes, V.~Vapnik, Support-vector networks, Machine learning 20~(3) (1995)
  273--297.

\bibitem{rosenblatt1958perceptron}
F.~Rosenblatt, The perceptron: a probabilistic model for information storage
  and organization in the brain., Psychological review 65~(6) (1958) 386.

\bibitem{ANNInit2016}
E.~Burnaev, P.~Erofeev, \href{https://doi.org/10.1134/S106422691606005X}{The
  influence of parameter initialization on the training time and accuracy of a
  nonlinear regression model}, Journal of Communications Technology and
  Electronics 61~(6) (2016) 646--660.
\newblock \href {http://dx.doi.org/10.1134/S106422691606005X}
  {\path{doi:10.1134/S106422691606005X}}.
\newline\urlprefix\url{https://doi.org/10.1134/S106422691606005X}

\bibitem{HDA2013}
M.~G. Belyaev, E.~V. Burnaev, Approximation of a multidimensional dependency
  based on a linear expansion in a dictionary of parametric functions,
  Informatics and its Applications 7~(3) (2013) 114--125.

\bibitem{Ensembles2013}
E.~V. Burnaev, P.~V. Prikhod'ko,
  \href{https://doi.org/10.1134/S0005117913100044}{On a method for constructing
  ensembles of regression models}, Automation and Remote Control 74~(10) (2013)
  1630--1644.
\newblock \href {http://dx.doi.org/10.1134/S0005117913100044}
  {\path{doi:10.1134/S0005117913100044}}.
\newline\urlprefix\url{https://doi.org/10.1134/S0005117913100044}

\bibitem{chollet2015keras}
F.~Chollet, et~al., Keras, \url{https://keras.io} (2015).

\bibitem{2016arXiv160502688short}
{Theano Development Team}, \href{http://arxiv.org/abs/1605.02688}{{Theano: A
  {Python} framework for fast computation of mathematical expressions}}, arXiv
  e-prints abs/1605.02688.
\newline\urlprefix\url{http://arxiv.org/abs/1605.02688}

\bibitem{paszke2017automatic}
A.~Paszke, S.~Gross, S.~Chintala, G.~Chanan, E.~Yang, Z.~DeVito, Z.~Lin,
  A.~Desmaison, L.~Antiga, A.~Lerer, Automatic differentiation in pytorch.

\bibitem{spesivtsev2018predictive}
P.~Spesivtsev, K.~Sinkov, I.~Sofronov, A.~Zimina, A.~Umnov, R.~Yarullin,
  D.~Vetrov, Predictive model for bottomhole pressure based on machine
  learning, Journal of Petroleum Science and Engineering 166 (2018) 825--841.

\bibitem{erofeev2019}
A.~Erofeev, D.~Orlov, A.~Ryzhov, D.~Koroteev, Prediction of porosity and
  permeability alteration based on machine learning algorithms, Transport in
  Porous Media (2019) 1--24.

\bibitem{kingma2014adam}
D.~P. Kingma, J.~Ba, Adam: A method for stochastic optimization, arXiv preprint
  arXiv:1412.6980.

\bibitem{minami1987liquid}
K.~Minami, J.~Brill, et~al., Liquid holdup in wet-gas pipelines, SPE Production
  Engineering 2~(01) (1987) 36--44.

\bibitem{abdul1996liquid}
G.~Abdul-Majeed, Liquid holdup in horizontal two-phase gas—liquid flow,
  Journal of Petroleum Science and Engineering 15~(2-4) (1996) 271--280.

\bibitem{mukherjee1979phd}
H.~Mukherjee, An experimental study of inclined two-phase flow, Ph.D. thesis,
  U. of Tulsa (1979).

\bibitem{eaton1966phd}
B.~Eaton, The prediction of flow patterns, liquid holdup and pressure losses
  occurring during continuous two-phase flow in horizontal pipelines, Ph.D.
  thesis, U. of Texas at Austin (1966).

\bibitem{beggs1973phd}
H.~Beggs, An experimental study of two-phase flow in inclined pipes, Ph.D.
  thesis, U. of Tulsa (1973).

\bibitem{AndritsosPhD}
N.~Andritsos, Effect of pipe diameter and liquid viscosity on horizontal
  stratified flow, Ph.D. thesis, U. of Illinois (1986).

\bibitem{Imbalance2015}
E.~Burnaev, P.~Erofeev, A.~Papanov,
  \href{https://doi.org/10.1117/12.2228523}{Influence of resampling on accuracy
  of imbalanced classification}, in: Proc. SPIE, Vol. 9875, 2015, pp. 9875 --
  9875 -- 5.
\newblock \href {http://dx.doi.org/10.1117/12.2228523}
  {\path{doi:10.1117/12.2228523}}.
\newline\urlprefix\url{https://doi.org/10.1117/12.2228523}

\bibitem{Imbalance2019}
D.~Smolyakov, A.~Korotin, P.~Erofeev, A.~Papanov, E.~Burnaev, Meta-learning for
  resampling recommendation systems, 2019.

\bibitem{ModelSelection2015}
E.~Burnaev, P.~Erofeev, D.~Smolyakov,
  \href{https://doi.org/10.1117/12.2228794}{Model selection for anomaly
  detection}, in: Proc. SPIE, Vol. 9875, 2015, pp. 9875 -- 9875 -- 6.
\newblock \href {http://dx.doi.org/10.1117/12.2228794}
  {\path{doi:10.1117/12.2228794}}.
\newline\urlprefix\url{https://doi.org/10.1117/12.2228794}

\bibitem{mandhane1974flow}
J.~Mandhane, G.~Gregory, K.~Aziz, A flow pattern map for gas—liquid flow in
  horizontal pipes, International Journal of Multiphase Flow 1~(4) (1974)
  537--553.

\bibitem{asheim1986mona}
H.~Asheim, et~al., Mona, an accurate two-phase well flow model based on phase
  slippage, SPE Production Engineering 1~(03) (1986) 221--230.

\bibitem{VovkConformal2014}
E.~Burnaev, V.~Vovk,
  \href{http://proceedings.mlr.press/v35/burnaev14.html}{Efficiency of
  conformalized ridge regression}, in: M.~F. Balcan, V.~Feldman, C.~Szepesvári
  (Eds.), Proceedings of The 27th Conference on Learning Theory, Vol.~35 of
  Proceedings of Machine Learning Research, PMLR, Barcelona, Spain, 2014, pp.
  605--622.
\newline\urlprefix\url{http://proceedings.mlr.press/v35/burnaev14.html}

\bibitem{ConformalKRR2016}
E.~Burnaev, I.~Nazarov, Conformalized kernel ridge regression, in: 2016 15th
  IEEE International Conference on Machine Learning and Applications (ICMLA),
  2016, pp. 45--52.
\newblock \href {http://dx.doi.org/10.1109/ICMLA.2016.0017}
  {\path{doi:10.1109/ICMLA.2016.0017}}.

\end{thebibliography}
\end{document}